\documentclass[aps, prb, twocolumn, superscriptaddress, amsmath,  tightenlines, longbibliography]{revtex4-1}

\usepackage{dcolumn}
\usepackage{graphicx}
\usepackage{mathrsfs}
\usepackage{subfigure}
\usepackage{booktabs}
\usepackage{amsmath}
\usepackage{physics}
\usepackage{dsfont}
\usepackage{amstext}
\usepackage{amssymb}
\usepackage{amsbsy}
\usepackage{bbm}
\usepackage{amsthm}
\usepackage{graphicx}
\usepackage{color}
\usepackage[colorlinks,citecolor=blue]{hyperref}

\usepackage{url}
\usepackage[colorlinks]{hyperref}
\hypersetup{%
	plainpages=true,
	breaklinks=true,       
	hypertexnames=false,  
	pageanchor=true,
	colorlinks=true,
	linkcolor={blue},
	citecolor={red},
	urlcolor={blue},
	anchorcolor={black}
}

 \makeatletter

\newcommand{\Rmnum}[1]{\expandafter\@slowromancap\romannumeral #1@}
\makeatother

\begin{document}

\title{Versatile Control of Nonlinear Topological States in Non-Hermitian Systems}
\author{Zhao-Fan Cai}
\affiliation{School of Physics and Optoelectronics, South China University of Technology,  Guangzhou 510640, China}
\author{Yu-Chun Wang}
\affiliation{School of Physics and Optoelectronics, South China University of Technology,  Guangzhou 510640, China}
\author{Yu-Ran Zhang}
\affiliation{School of Physics and Optoelectronics, South China University of Technology,  Guangzhou 510640, China}
\author{Tao Liu}
\email[E-mail: ]{liutao0716@scut.edu.cn}
\affiliation{School of Physics and Optoelectronics, South China University of Technology,  Guangzhou 510640, China}
\author{Franco Nori}
\affiliation{Center for Quantum Computing, RIKEN, Wakoshi, Saitama 351-0198, Japan}
\affiliation{Department of Physics, University of Michigan, Ann Arbor, Michigan 48109-1040, USA}

\date{{\small \today}}


\begin{abstract}
	The non-Hermitian skin effect (NHSE) and nonlinearity can both delocalize topological modes (TMs) from the interface. However, the NHSE requires precise parameter tuning, while the nonlinearity in Hermitian systems results in partial delocalization with limited mode capacity. To overcome these limitations, we propose a non-Hermitian nonlinear topological interface model that integrates Hermitian and non-Hermitian lattices with nonreciprocal hopping and nonlinearity. This system enables the complete delocalization of TMs across the entire lattice without fine-tuning, while allowing precise control over the wavefunction profile and spatial distribution through the intrinsic configuration and intensity of the nonlinearity.  Using the spectral localizer, we demonstrate the topological protection and robustness of these extended non-Hermitian TMs against disorder. Furthermore, we show that under external pumping, localized excitations evolve into predefined profiles and generate long-range patterns, an effect unattainable in Hermitian systems. These findings reveal how the interplay of nonlinearity and NHSE shapes topological states, paving the way for compact topological devices.
\end{abstract}
\maketitle

\vspace{.3cm}
\noindent {\large \textbf{Introduction}}

\noindent Over the past decade, topological phases have emerged as one of the most rapidly advancing research areas, garnering significant attention across diverse fields such as condensed matter \cite{RevModPhys.82.3045,RevModPhys.83.1057,RevModPhys.88.035005}, photonics \cite{Lu2014,RevModPhys.91.015006,Leefmans2022}, and electrical circuits systems \cite{Imhof2018,Lee2018,Yang2024}.  A hallmark feature of topological phases is the presence of topological modes (TMs), typically confined to system boundaries or interfaces, and governed by the bulk-boundary correspondence \cite{RevModPhys.82.3045}. These TMs decay exponentially into the bulk  and remain resilient to disorder and perturbations, granting them inherent robustness. However, since the existence of TMs depends on bulk-band topology, their reliance on bulky materials with large lattice structures poses a significant challenge for scalability and integration into compact systems, hindering practical applications and the broader advancement of topological technologies.

A recent study suggests that incorporating non-Hermiticity into the system can overcome the boundary-localized nature of TMs, with the non-Hermitian skin effect (NHSE) facilitating their delocalization \cite{PhysRevLett.125.206402, Wang2022, PhysRevB.103.195414}. The NHSE is characterized by the collapse of bulk-band eigenstates into localized boundary modes, and its discovery has opened new avenues for exploring exotic physics that have no Hermitian counterparts in non-Hermitian systems \cite{PhysRevLett.118.040401,Gao2015,Monifi2016,ZhangJ2018,  PhysRevLett.118.045701,arXiv:1802.07964,Peng2014b,El-Ganainy2018,ShunyuYao2018,PhysRevLett.125.126402,PhysRevLett.123.066404,YaoarXiv:1804.04672,PhysRevLett.121.026808,PhysRevLett.122.076801,PhysRevLett.123.170401, PhysRevLett.123.206404,PhysRevLett.123.066405,PhysRevLett.123.206404, Li2020,PhysRevLett.124.086801, PhysRevLett.125.186802, PhysRevB.100.054105, Zhao2019,PhysRevX.9.041015, PhysRevB.102.235151,Bliokh2019,   PhysRevAL061701,  PhysRevLett.127.196801, RevModPhys.93.015005,   Parto2023, Ren2022,PhysRevX.13.021007,PhysRevLett.131.116601,PhysRevA.109.063329,PhysRevX.14.021011,arXiv:2408.12451,PhysRevLett.131.036402,PhysRevResearch.7.L012068,PhysRevLett.132.050402,arXiv:2404.16774,Leefmans2024,PhysRevLett.133.136602,arXiv:2311.03777,PhysRevB.111.205418}. By  harnessing the NHSE,  the wavefunction  of the TM can transition from localized to delocalized states when the system parameters satisfy a critical condition  \cite{Wang2022}. 
However, the requirement for such precise parameter tuning imposes practical limitations on the implementation of extended TMs and increases fabrication complexity. To overcome this challenge, we propose a non-Hermitian and nonlinear approach that circumvents these constraints, offering a more robust and scalable solution.

Highly controllable nonlinearity has emerged as a powerful and versatile tool for manipulating next-generation topological devices \cite{Smirnova2020}. They enable a broad range of novel topological phenomena, including the formation of topological soliton states \cite{PhysRevLett.111.243905, PhysRevLett.117.143901, PhysRevX.11.041057, Mukherjee2020, PhysRevLett.127.184101, PhysRevB.110.L180302}, the nonlinearity-driven topological phase transitions \cite{PhysRevLett.123.053902, PhysRevB.102.115411, Sone2024, Sone2025}, and the active control of TMs when combined with non-Hermiticity \cite{Xia2021, Dai2023}. Recently,  it was demonstrated that the nonlinearity can partially delocalize topological zero modes (TZMs), which are originally confined to the interface between a nonlinear and  a linear Hermitian Su-Schrieffer-Heeger (SSH) chains. This effect causes the TZMs to spread throughout the entire nonlinear chain \cite{PhysRevLett.133.116602}. Moreover, the wavefunction profiles of TZMs can be tailored to form arbitrary plateaus,  opening new avenues for topological state manipulation \cite{PhysRevLett.133.116602}.

Despite significant progress, a fundamental question remains unresolved: Can the synergy of nonlinearity, non-Hermiticity, and topology enable fully delocalized and reconfigurable TMs across an entire lattice without strict parameter constraints? Addressing this challenge is crucial for overcoming the limitations of current topological systems that rely on bulky materials, expanding their functional scope, and enabling compact and reconfigurable device architectures. Moreover, achieving fully extended states beyond the continuum could significantly enhance a wide range of topological applications  \cite{Wang2022}.

In this work, we explore the intricate interplay between nonlinearity, NHSE, and topology to achieve the delocalization and precise design of TMs. Initially confined to the interface between Hermitian and non-Hermitian lattices, these modes exhibit unusual behavior under the influence of nonlinearity. When the nonlinearity is introduced exclusively in the Hermitian section of the topological interface model, the TM extends across the entire Hermitian part under weak nonreciprocal hopping. As the system parameters approach a critical condition, the mode seamlessly expands over both the Hermitian and non-Hermitian regions. In contrast, strong nonreciprocity localizes the mode at the boundary due to NHSE. When the nonlinearity is applied to both the Hermitian and non-Hermitian regions, the TM becomes fully delocalized across the entire lattice without requiring precise parameter tuning, significantly enhancing its versatility and robustness. Moreover, this approach allows for arbitrary shaping of wavefunction profiles, enabling customized configurations across the entire lattice. The topological protection of these extended non-Hermitian TMs is rigorously verified using the real-space spectral localizer. Additionally, we explore their dynamical stability under external pumping, demonstrating how an initially localized excitation evolves into a predefined wavefunction profile. Furthermore, the nonlinear non-Hermitian model enables the excitation of long-range patterns, an effect unattainable in purely Hermitian systems. 

\begin{figure}[!b]
	\centering
	\includegraphics[width=8.7cm]{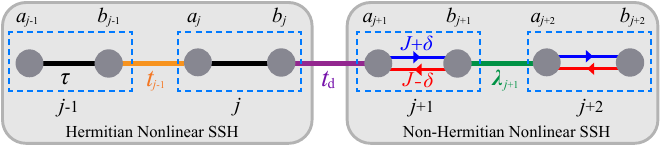}
	\caption{\textbf{Non-Hermitian nonlinear topological interface lattice}. Tight-binding representation of a one-dimensional topological interface model consisting of a Hermitian nonlinear SSH chain (left) and a non-Hermitian nonlinear  SSH chain lattice (right). For the Hermitian  chain, $\tau$ is  the intracell hopping strength (black link), and $t_{j} = \tilde{t}_{j} + \alpha (\abs{a_{j+1}}^2 + \abs{b_{j}}^2)$ is the intensity-dependent nonlinear intercell hopping strength (orange link) with $\alpha$ being the Kerr nonlinear coefficient. For the non-Hermitian  chain, $J\pm \delta$ denote  nonreciprocal intracell hopping amplitudes (blue and red links), and $\lambda_{j} = \tilde{\lambda}_j + \beta (\abs{a_{j+1}}^2 + \abs{b_{j}}^2)$ is the nonlinear intercell hopping strength (green link) with   Kerr nonlinear coefficient $\beta$. Also, $t_\text{d}$ represents the inter-chain coupling strength (violet link). }\label{Fig1}
\end{figure}

\begin{figure*}[!tb]
	\centering
	\includegraphics[width=17.8cm]{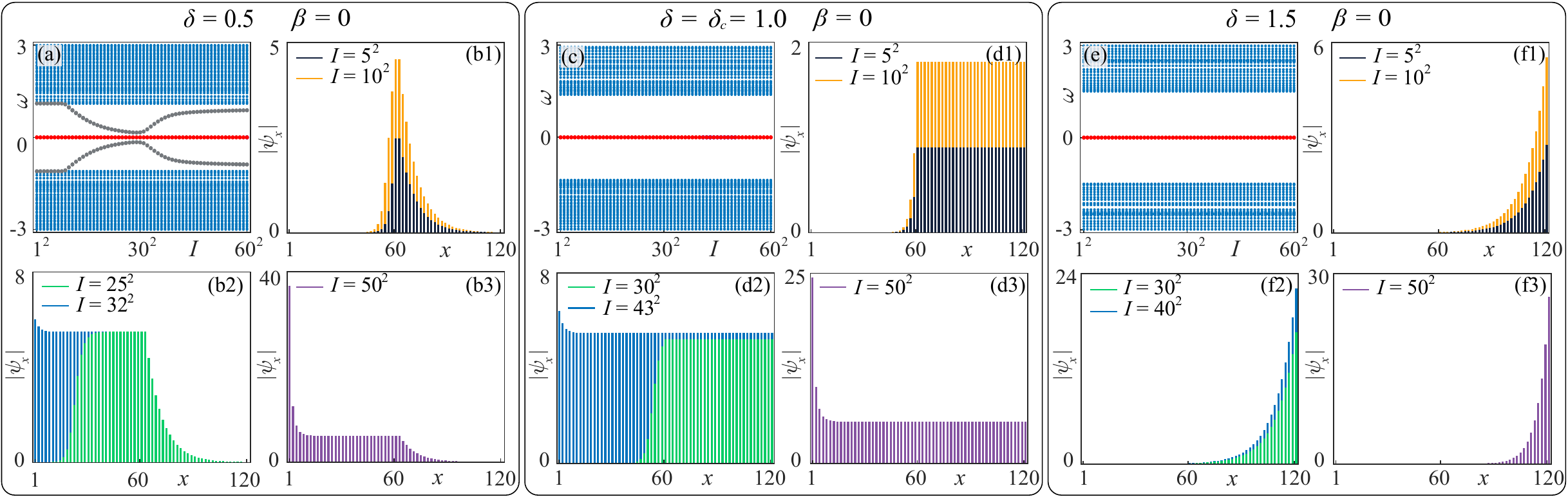}
	\caption{\textbf{Complete delocalization of a topological zero modes (TZMs) under a critical condition}. Eigenfrequency spectrum $\omega$ versus the squared amplitudes $I= \sum_{j} (\abs{a_j}^2 + \abs{b_j}^2)$ for Kerr nonlinear coefficient  $\beta = 0$ with (a) $\delta=0.5$, (c) $\delta=\delta_\text{c}=1.0$, and (e) $\delta=1.5$, where red dots indicate the TZMs. In (a),  in-gap non-zero  modes (gray dots) are shown. The corresponding spatial distributions $\abs{\psi_x}$  of TZMs for different $I$ are shown in (b1-b3), (d1-d3) and (f1-f3), respectively.   Other parameters used here are $J=1.5$, $\tilde{t}_j = 1.0$, $\tau=t_\text{d}=\tilde{\lambda}_j =2.5$,   $\alpha=0.05$,  $N=31$ and $L=121$. }\label{Fig2}
\end{figure*}

\vspace{.3cm}
\noindent {\large \textbf{Results and Discussion}}

\noindent \textbf{Model}

\noindent We consider a  one-dimensional (1D) topological interface model consisting of a  Hermitian nonlinear SSH   chain  and a non-Hermitian nonlinear SSH   chain, as shown in Fig.~\ref{Fig1}. Both SSH chains contain two sublattices $a$ and $b$, and their  state amplitudes are labeled by $a_j$ and $b_j$ on the $j$-th unit cell with $j \leq N$ for the Hermitian chain, and  $j > N $ for the non-Hermitian chain. The real-space eigenvector $\ket{\psi}=(\cdots,a_j,b_j,\cdots)^T$ of this hybrid system is captured by the nonlinear Schr\"{o}dinger equation $\hat{\mathcal{H}} \ket{\psi} = \omega \ket{\psi}$, where  $\omega$ is  eigenfrequency, and  the   Hamiltonian $\hat{\mathcal{H}}$ is written as
\begin{align}\label{Hamil}
	\hat{\mathcal{H}} =&  \sum_{j\leq N} \left(\tau \ket{a_j} \bra{b_j} + t_{j-1} \ket{a_{j}} \bra{b_{j-1}} + \text{H.c.}\right)  \nonumber \notag \\
	& + \sum_{j>N} \left[(J-\delta) \ket{a_j} \bra{b_j} + (J+\delta) \ket{b_j} \bra{a_j} \right] \nonumber \notag \\
	& + \sum_{j>N}  \left( \lambda_j \ket{a_{j+1}} \bra{b_j} + \text{H.c.} \right) \nonumber \notag \\
	& + t_\text{d} \left(\ket{a_{N+1}} \bra{b_N} + \text{H.c.}\right).
\end{align}

In Eq.~(\ref{Hamil}),  $\tau$ is the intracell hopping strength in the  Hermitian chain (see Fig.~\ref{Fig1}), and $t_{j} = \tilde{t}_j + \alpha (\abs{a_{j+1}}^2 + \abs{b_{j}}^2)$ 	is the intensity-dependent nonlinear intercell hopping strength, with $\alpha$ being the Kerr nonlinear coefficient. $J\pm \delta$ denote the nonreciprocal intracell hopping amplitudes in the non-Hermitian chain, $\lambda_j = \tilde{\lambda}_j + \beta (\abs{a_{j+1}}^2 + \abs{b_{j}}^2)$ is the nonlinear intercell hopping strength, with $\beta$ being the Kerr nonlinear coefficient, and $t_\text{d}$ represents the inter-chain coupling strength.  Such non-Hermitian nonlinear lattices can be feasibly implemented in various experimental platforms, including photonic systems \cite{PhysRevA.100.063830,Sone2025} and electronic  circuits \cite{Hadad2018,arXiv:2403.10590,arXiv:2505.09179} [See detailed discussion in Supplementary Note 1].    Unless otherwise specified, we assume that the Hermitian SSH chain is in the topologically trivial regime, while the non-Hermitian chain is topologically nontrivial.

\vspace{.3cm}
\noindent \textbf{Nonlinear  morphing of TZMs in a critical condition}

\noindent Without nonlinearity ($\alpha=\beta=0$), a TZM can be delocalized from the interface, occupying  only the non-Hermitian SSH chain of the interface model at a critical value\cite{Wang2022} with  $\delta_\text{c}=\lambda-J$, where we label $\tilde{\lambda}_j=\lambda$. Here, we demonstrate that the TZM can occupy the entire Hermitian and non-Hermitian chains when the nonlinearity is only applied to the Hermitian chain ($\alpha \neq 0$, $\beta=0$). We solve self-consistently the nonlinear Schr\"{o}dinger equation   for different squared amplitudes $I = \sum_{j} (\abs{a_j}^2 + \abs{b_j}^2)$. 

Figure \ref{Fig2} shows eigenfrequency spectrum $\omega$ versus  $I$, and the corresponding spatial distributions $\abs{\psi_x} = \{\abs{a}, \abs{b}\} $ ($x$ denotes the lattice site) of the TZMs [see red dots in Fig.~\ref{Fig2}(a,c,e)] for different  $\delta$ with $\alpha = 0.05$. Note that, in addition to TZMs, the nonlinearity can induce in-gap non-zero  modes  [see gray dots in Fig.~\ref{Fig2}(a)]. This work focuses exclusively on TZMs, with detailed discussions of the non-zero in-gap modes provided in Supplementary Note 2. 

For weak nonreciprocal hopping with $\delta=0.5 < \delta_\text{c}$,   a TZM is localized at the interface  between the topologically trivial Hermitian nonlinear chain and the nontrivial non-Hermitian linear chain for small $I$, as shown in Fig.~\ref{Fig2}(b1). As $I$ increases, the TZM gradually delocalizes towards the Hermitian chain, and  eventually spreads uniformly across the entire Hermitian chain at large values of $I$, e.g., $I=32^2$ [see Fig.~\ref{Fig2}(b2)]. Furthermore, as $I$ increases further, the TZM becomes compressed toward the left boundary while maintaining a plateau in the  bulk region of the Hermitian lattice at sufficiently  large values of $I$, e.g., $I=50^2$ [see Fig.~\ref{Fig2}(b3)]. Note that the same nonlinearity-induced delocalization of TZMs is also observed\cite{PhysRevLett.133.116602} in the Hermitian case with $\delta=0$.

Distinctive behavior arises at the critical point   $\delta=\delta_\text{c}$. For the weak nonlinearity, the TZM occupies the whole non-Hermitian chain due to the competition between NHSE localization and topological localization of the in-gap interface state \cite{Wang2022}, as shown in Fig.~\ref{Fig2}(d1). However, as $I$ increases, the TZM gradually spreads to uniformly occupy the entire Hermitian and non-Hermitian lattices due to the interplay of the nonlinearity, NHSE and topology  at large values of $I$, e.g., $I=43^2$ [see Fig.~\ref{Fig2}(d2)]. Furthermore, for significantly large values of $I$, e.g., $I=50^2$ in Fig.~\ref{Fig2}(d3), the TZM is compressed toward the left boundary while maintaining a plateau throughout the entire  bulk region of two chains. In the case of strong nonreciprocal hopping [e.g., $\abs{\psi_x}$ for $\delta=1.5$ in Fig.~\ref{Fig2}(f1-f3)], the NHSE dominates the nonlinear effects, and the TZM is  localized at the right boundary even for strong nonlinearity. These results show that the  interplay of nonlinearity, nonreciprocal hopping, and topology determines the morphing of TZM wavefunctions. Note that the delocalized TZM remains robust against disorder (see details in Supplementary Note 3).

\begin{figure*}[!tb]
	\centering
	\includegraphics[width=17.8cm]{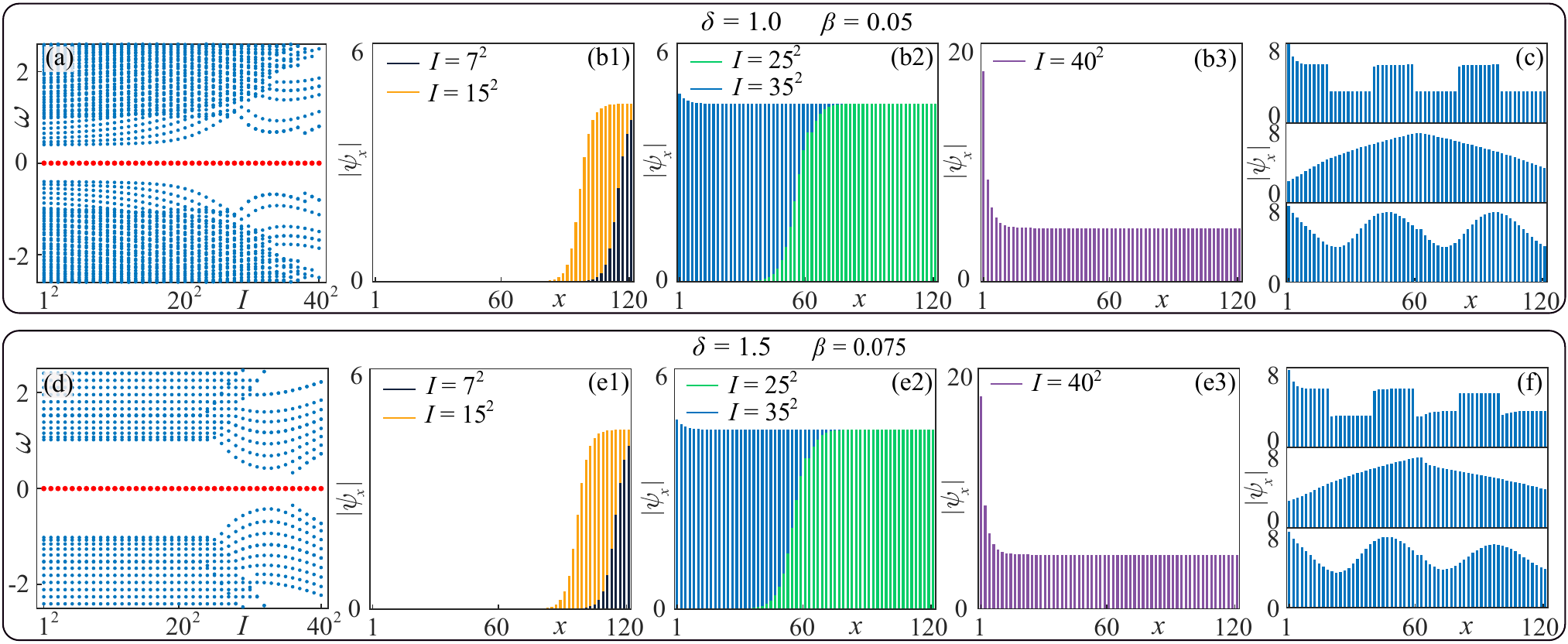}
	\caption{\textbf{Delocalization and manipulation of topological zero modes (TZMs) without fine parameter tuning}. Eigenfrequency $\omega$ versus the squared amplitudes $I= \sum_{j} (\abs{a_j}^2 + \abs{b_j}^2)$ (a) for $\delta = 1$ and $\beta = 0.05$,  and (d) for $\delta = 1.5$ and $\beta = 0.075$, where red dots indicate the TZMs.  The corresponding spatial distributions  $\abs{\psi_x}$  of TZMs for different $I$, with $\tilde{t}_j = \tilde{\lambda}_j = 1.5$, are shown in (b1-b3) and (e1-e3), respectively. 	Square,  isosceles triangle and  cosine profiles  of  TZMs (c) for $\delta = 1$ and $\beta = 0.05$,  and (f) for $\delta = 1.5$ and $\beta = 0.075$ by designing $\tilde{t}_j$ and $\tilde{\lambda}_j$, where the corresponding distributions $\tilde{t}_j$ and $\tilde{\lambda}_j$ are provided in Supplementary Figure S5. Other parameters used here are  $J=1.5$, $\alpha=0.05$, $\tau=t_\text{d}=2.5$, $N=31$, and $L=121$. }\label{Fig3}
\end{figure*}

\begin{figure*}[!tb]
	\centering
	\includegraphics[width=16cm]{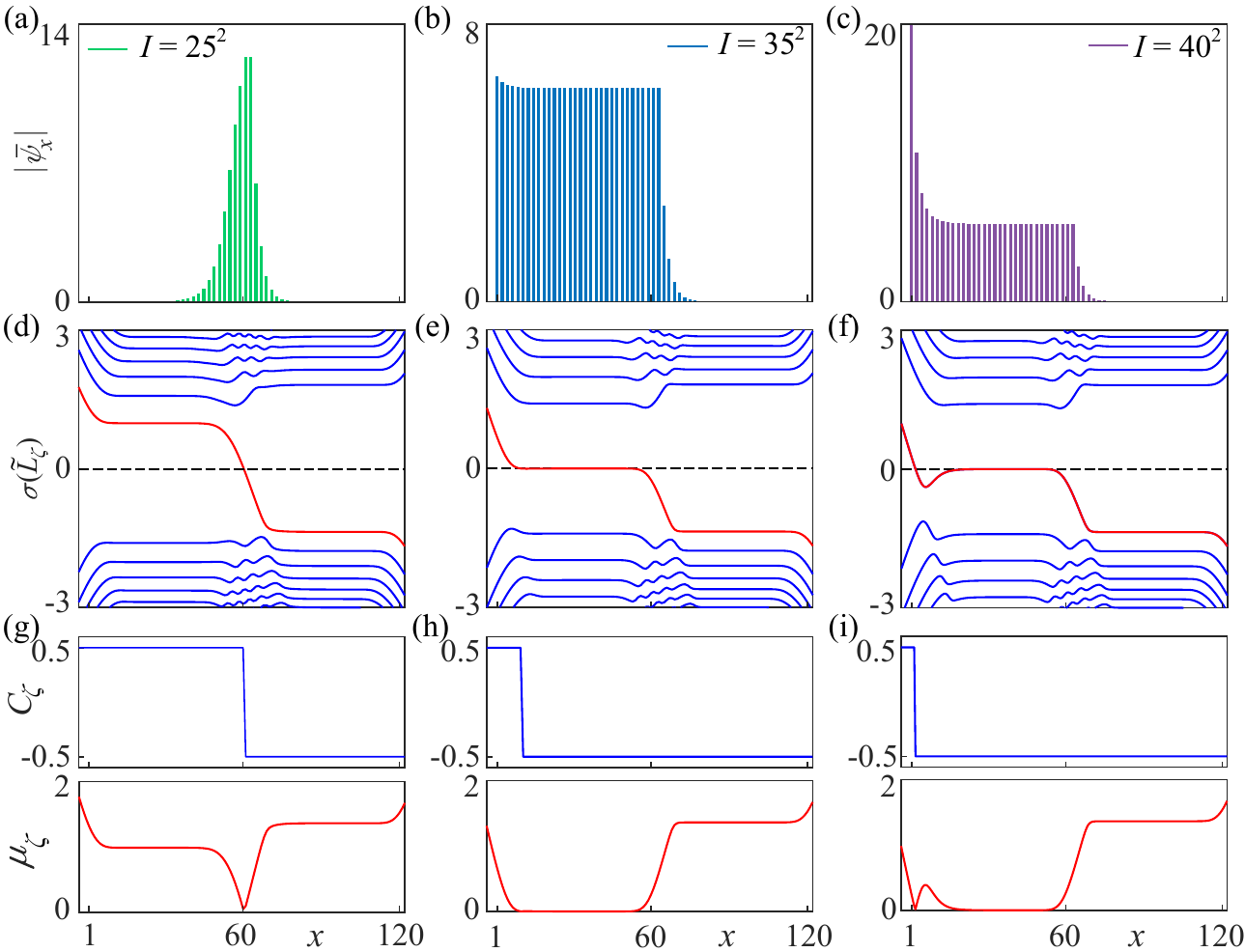}
	\caption{\textbf{Spectral localizer}. (a-c) Spatial distributions $\abs{\bar{\psi}_x} = \abs{\hat{S} \psi_x}$ of topological zero modes.  (d-f) Eigenvalues of $\tilde{L}_{\zeta}$, labeled by $\sigma(\tilde{L}_\zeta)$, versus  $x$ for different $I = \sum_{j} (\abs{a_j}^2 + \abs{b_j}^2)$, with $\eta=0.2$ and $\bar{\omega}=0$. (g-i) Site-resolved topological invariant $C_\zeta =  \text{Sig} (\tilde{L}_\zeta)/2$, and local band gap $\mu_\zeta = \abs{\sigma_\text{min}(\tilde{L}_\zeta ) }$ for different $I$. The distributions $\tilde{t}_j$ and $\tilde{\lambda}_j$, along with the other parameters,  are chosen to be the same as in the case $\delta=1$ and $\beta=0.05$ shown in Fig.~\ref{Fig3}.}\label{Fig4}
\end{figure*}

\vspace{.3cm}
\noindent \textbf{Nonlinearity-enabled control of TZMs}

\noindent The TZM can occupy the entire  lattice when $\delta=\delta_\text{c}$ with $\beta=0$, which can benefit a wide variety of topological applications \cite{Wang2022}. However, the necessary condition with  $\delta_\text{c}=\lambda-J$ in the non-Hermitian linear  chain restricts its tunability. To overcome this limitation, we introduce the nonlinearity into the non-Hermitian chain, i.e., $\beta \ne 0$ for $\hat{\mathcal{H}}$ in Eq.~(\ref{Hamil}), and focus on parameters that deviate from the linear critical condition  with $(J + \delta) > \lambda$. Our analysis considers the scenario where the linear non-Hermitian SSH chain resides in the topologically nontrivial regime, characterized by  $J \in [-\sqrt{\delta^2 + \lambda^2},~\sqrt{\delta^2 + \lambda^2}]$ for $\abs{\lambda} \geq \abs{\delta}$.

Figure \ref{Fig3} shows the  eigenfrequency  $\omega$ versus  $I$, and the corresponding spatial distributions $\abs{\psi_x}$  of TZMs [red dots in Fig.~\ref{Fig3}(a,d)] for  $\delta=1$ with $\beta = 0.05$ [see Fig.~\ref{Fig3}(b1-b3)], and $\delta=1.5$ with $\beta = 0.075$ [see Fig.~\ref{Fig3}(e1-e3)]. When  $\delta=1$, in contrast to the case of the non-Hermitian linear chain [see Fig.~\ref{Fig2}(d1)], the TZM is initially localized at the right boundary due to the NHSE, and then gradually spreads from the boundary as $I$ increases [see Fig.~\ref{Fig3}(b1)]. For large values of $I$, the TZM occupies the entire lattice [see Fig.~\ref{Fig3}(b2-b3)], forming a plateau. The most notable finding is that the delocalization of TZMs, accompanied by the occupation of the entire lattice, occurs without requiring $\delta = 1$ for $\beta \neq 0$. For example, even with unidirectional hopping at $\delta=1.5$, a uniform distribution of TZMs across the entire lattice is observed [see Fig.~\ref{Fig3}(e2)], which does not occur for $\beta=0$. Further details on how $\delta$ and $\beta$ influence the plateau behavior of TZMs are provided in Supplementary Note 4. 

Along the entire lattice, arbitrary wavefunction profiles of TZMs can be achieved by designing site-dependent hopping parameters, $\tilde{t}_j$ and $\tilde{\lambda}_j$ [see Supplementary Note 5(A)], without requiring a linear critical condition. Figures \ref{Fig3}(c) and (f) illustrate square, isosceles triangle, and cosine profiles of TZMs    for $\delta = 1$ and  $\delta = 1.5$, respectively. Notably, even when the system parameters   deviate from the linear critical condition, ideal wavefunction profiles can still be achieved by adjusting $\beta$, as demonstrated by comparing Figs.~\ref{Fig3}(c) and \ref{Fig3}(f). Further details regarding  nonlinearity-driven control of TZMs and disorder effects  are provided in Supplementary Note 5. 

\begin{figure*}[!tb]
	\centering
	\includegraphics[width=18cm]{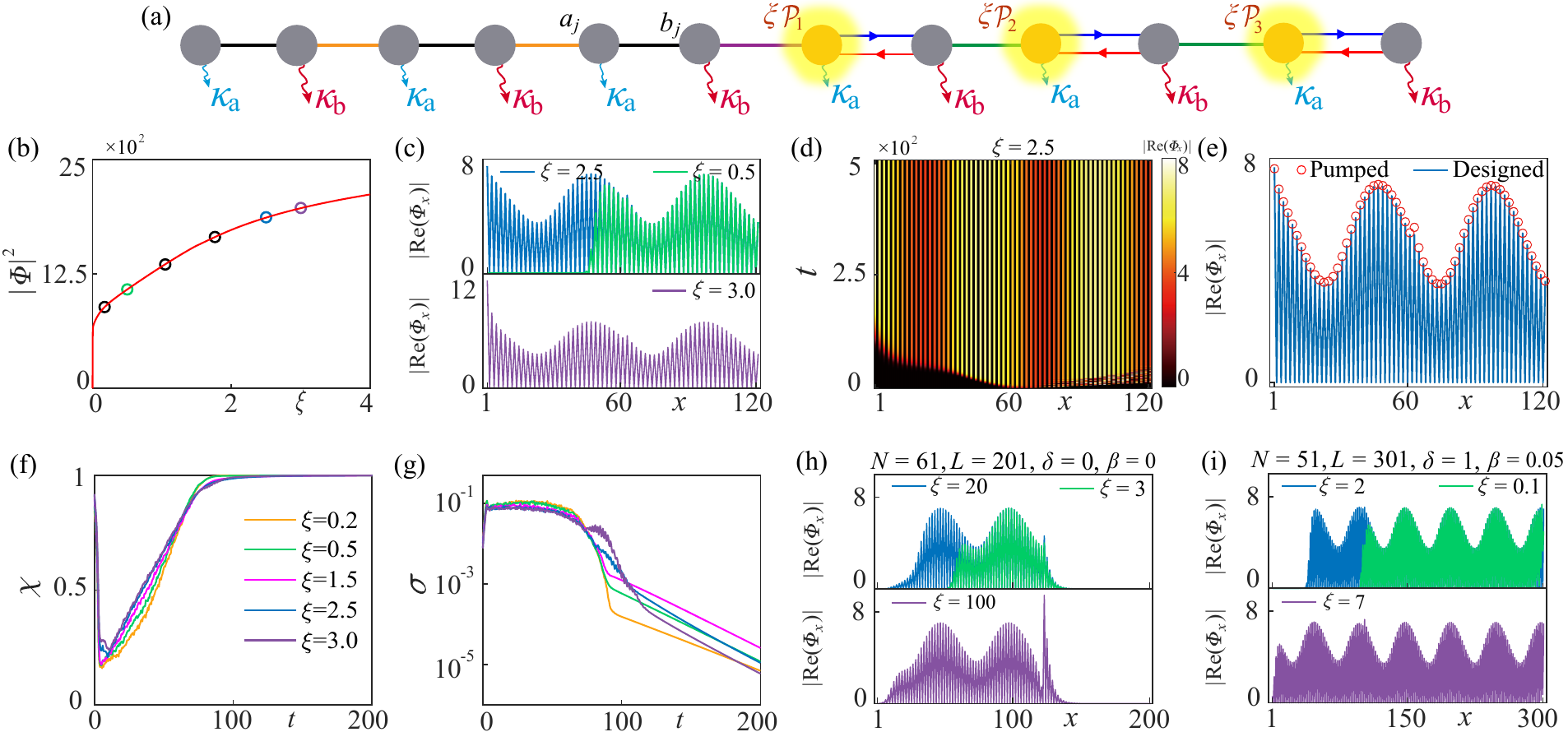}
	\caption{\textbf{Dynamical evolution and external excitation of tailored topological zero modes (TZMs)}. (a) Schematic showing external pumping applied to the $a$-sites of the non-Hermitian chain, highlighted by yellow spots. The intensity distributions of the pumping sources are represented by $\xi \mathcal{P}_m$, with $\xi$ being the pumping strength. The cyan and red wave arrows indicate staggered onsite losses in the two sublattices, labeled  $\kappa_\text{a}$ and $\kappa_\text{b}$, and different colored lines connecting the chain sites correspond to different hopping terms in Fig.~\ref{Fig1}. (b) Intensity $\abs{\Phi}^2$ of the evolved steady state versus $\xi$. The circles represent  results from the evolution equation  (\ref{SchrodingerEquationDynamics}), which closely match the results   (red curve)  from   self-consistent nonlinear equations  in the Supplementary Equation (S10). (c) Wavefunction profile $\abs{\textrm{Re}(\Phi_x)}$ of the evolved steady states for different $\xi$,   corresponding to colored circles in (b). (d) Time- and space-resolved $\abs{\textrm{Re}(\Phi_x)}$ for $\xi = 2.5$,  with the steady-state result  highlighted by red circles in   (e).    In (e), the steady-state wavefunction profile (red circles) closely matches the designed one (blue line) of the TZM. Evolution of (f) the similarity function $\chi(t)$ and (g) the corresponding standard deviations $\sigma(t)$ for different $\xi$ with $200$ independent realizations. (h,i)  $\abs{\textrm{Re}(\Phi_x)}$ illustrating the excitation of a long-range pattern with a large-size lattice in the (h) Hermitian  and (i) non-Hermitian interface models.	The parameters    are the same as the case of $\delta=1$ and $\beta=0.05$ in Fig.~\ref{Fig3}, with $\tilde{\omega}=0$, $\kappa_\text{a}=0.01$ and $\kappa_\text{b}=0.5$. }\label{Fig5}
\end{figure*}

\vspace{.3cm}
\noindent \textbf{Topological origin of zero modes}

\noindent Conventional topological invariants are typically characterized by the system's band structure and associated Bloch eigenstates, and are regarded as global properties of the system. However, nonlinear effects in our model are inherently local, and strong nonlinearity breaks translation symmetry, rendering topological invariants in momentum space ill-defined. To verify the topological origin of the zero modes in our nonlinear non-Hermitian system, we utilize a spectral localizer \cite{PhysRevLett.131.213801,PhysRevB.108.195142,Cheng2023,PhysRevB.108.035107,PhysRevB.109.035425,Ochkan2024}, which is applicable to systems lacking translation symmetry. In order to employ the spectral localizer to our 1D non-Hermitian system (see details in  Methods), we map the non-Hermitian Hamiltonian $\hat{\mathcal{H}}$ to the Hermitian one $\hat{\mathcal{H}}_\text{S}$ via a similarity transformation $\hat{S}$, i.e., $\hat{\mathcal{H}}_\text{S} = \hat{S} \hat{\mathcal{H}} \hat{S}^{-1}$,  with its eigenvector satisfying $\ket{\bar{\psi}} = \hat{S} \ket{\psi}$. The spectral localizer of a 1D system at any choice of location $x$ and frequency $\bar{\omega}$  is  written as \cite{Cheng2023}
\begin{align}\label{localizer}
	L_{\zeta\equiv(x,\bar{\omega})}(\hat{X},\hat{\mathcal{H}}_\text{S}) = \eta(\hat{X}-x\mathbf{I}) \otimes \Gamma_x + (\hat{\mathcal{H}}_\text{S}-\bar{\omega}) \otimes \Gamma_y,
\end{align}
where $\Gamma_x$ and $\Gamma_y$ are Pauli matrices, $\mathbf{I}$ is an identity matrix, $\eta$ is a tuning parameter which ensures $\hat{X}$ and $\hat{\mathcal{H}}_\text{S}$ have compatible units, and $\hat{X}$ is a diagonal matrix whose entries correspond to the coordinates of each lattice site.  When the system preserves chiral symmetry, the spectral localizer can be written in a reduced form as\cite{Cheng2023}
\begin{align}\label{localizerChiral}
	\tilde{L}_{\zeta\equiv(x,\bar{\omega})}(\hat{X},\hat{\mathcal{H}}_\text{S}) = \eta(\hat{X}-x\mathbf{I})\hat{\Pi} + \hat{\mathcal{H}}_\text{S} -i \bar{\omega} \hat{\Pi},
\end{align}
with $\hat{\Pi}$ being the system's chiral operator. The eigenvalues of $\tilde{L}_{\zeta}$ is labeled by $\sigma(\tilde{L}_\zeta)$, and the local band gap is given by the smallest value as $\mu_\zeta = \abs{\sigma_\text{min}(\tilde{L}_\zeta ) }$. Furthermore, the local topological invariant is written by \cite{Cheng2023}
\begin{align}\label{localinvariant}
	C_\zeta = \frac{1}{2} \text{Sig} (\tilde{L}_\zeta(\hat{X},\hat{\mathcal{H}}_\text{S})), 
\end{align}
where $\text{Sig}$  is the signature of a matrix, i.e., its number of positive eigenvalues minus its number of negative ones.

Figure \ref{Fig4} shows spatial distributions $\abs{\bar{\psi}_x} = \abs{\hat{S} \psi_x}$ of TZMs after a similarity transformation $\hat{S}$, site-resolved $\sigma(\tilde{L}_\zeta)$, $C_\zeta$, and $\mu_\zeta$   for different $I$ at $\bar{\omega}=0$, with $\delta=1$ and $\beta=0.05$. Specifically,   $\sigma(\tilde{L}_\zeta)$, $C_\zeta$, and $\mu_\zeta$ exhibit intensity dependence. Whenever  $\sigma(\tilde{L}_\zeta)$ crosses zero [see red curves in Fig.~\ref{Fig4}(d-f)], the value of $C_\zeta$ changes accordingly  [see blue curves in Fig.~\ref{Fig4}(g-i)]. Simultaneously, the local gap of $\mu_\zeta$ closes [see red curves in Fig.~\ref{Fig4}(g-i)]. The fact that a zero value of  $\sigma(\tilde{L}_\zeta)$ at $\zeta = \{x_0, 0\}$ signifies a zero-frequency mode that is  localized near at $x_0$,  implies that the change  in $C_\zeta$ at $\zeta = \{x_0, 0\}$  reflects the bulk-boundary correspondence. Furthermore, the regions where the local band gap $\mu_\zeta$ is closest to zero differ for each intensity, signifying the emergence of extended TZMs. Further details on the spectral localizer for different wavefunction profiles  are provided in Supplementary Note 6.

This framework allows us to rigorously assess the topological protection of the TZMs. Specifically, the robustness of a TZM can be guaranteed as long as any perturbation to the system remains below the local band gap. This condition is expressed by\cite{PhysRevLett.133.116602}
\begin{align}\label{topoprotect}
	\abs{\abs{\Delta \hat{\mathcal{H}}_\text{S}(W)}} < \mu_\zeta^\text{max},
\end{align}
where $\abs{\abs{\Delta \hat{\mathcal{H}}_\text{S}(W)}}$ is the largest singular value of $\Delta\hat{\mathcal{H}}_\text{S}(W) = \hat{\mathcal{H}}_\text{S}(W) - \hat{\mathcal{H}}_\text{S}$, with $\hat{\mathcal{H}}_\text{S}(W) = \hat{\mathcal{H}}_\text{S} + W \delta\hat{\mathcal{H}}_\text{S}$ representing the perturbed  nonlinear Hamiltonian with perturbation strength $W$, and $\mu_\zeta^\text{max}=\max [\mu_\zeta]$ is the maximum value of $\mu_\zeta$ within the topological region. When this condition is satisfied, a stable TZM exists with topological protection. Further detailed discussion on disorder robustness of TZMs for different wavefunction profiles are provided in Supplementary Note 5(B).

\vspace{.3cm}
\noindent \textbf{Dynamical evolution under external pumping}

\noindent Unlike conventional linear topological models, nonlinear models can exhibit distinctive dynamical properties that depend on how intensity levels are reached, enabling intrinsic control on TZMs through external pumping  \cite{PhysRevLett.133.116602}. Here, we investigate the dynamical evolution under an external pumping scheme, as illustrated in Fig.~\ref{Fig5}(a), with the evolution governed by the following equation
\begin{align}\label{SchrodingerEquationDynamics}
	\frac{\partial \ket{\Phi}}{\partial t} = -i\left(\hat{\mathcal{H}} + \hat{\mathcal{H}}_\text{loss}\right)\ket{\Phi} + \xi \ket{\mathcal{P}} e^{-i\tilde{\omega} t},
\end{align}
where $\hat{\mathcal{H}}_\text{loss} = \sum_{j} (-i \kappa_\text{a} \ket{a_j} \bra{a_j} - i \kappa_\text{b} \ket{b_j}  \bra{b_j})$, denotes onsite losses in the two sublattices, which contributes to the stabilizing excitation. The pumping sources $\ket{\mathcal{P}} \equiv (0,\cdots,\mathcal{P}_m,\cdots)^T$ ($m=1,2,\cdots,(L-1)/2-N$) are only applied to the $a$-sites of the non-Hermitian chain  [see Fig.~\ref{Fig5}(a)], with the pumping frequency denoted by $\tilde{\omega}$, and the pumping strength $\xi$.  We consider a single external pumping source with distribution  $\mathcal{P}_m=\delta_{m,1}$ for $\tilde{\omega}=0$, and a complementary discussion on the alternative distribution for  $\mathcal{P}_m$ is provided in Supplementary Note 7(A).

Figure \ref{Fig5}(b) plots the intensity $\abs{\Phi}^2$ of the evolved steady state versus $\xi$, with the wavefunction profile $\abs{\textrm{Re}(\Phi_x)}$ for different $\xi$ shown in Fig.~\ref{Fig5}(c). As the pumping strength $\xi$ increases,  an initially localized waveform begins to spread and gradually occupies the non-Hermitian chain [see green line in Fig.~\ref{Fig5}(c)]. Eventually, this expansion allows the waveform to fill the entire lattice of both the Hermitian and non-Hermitian chains,  aligning with the designed profile  (indicated by the blue line).   Upon further increasing $\xi$, the steady-state intensity becomes concentrated toward the left boundary while maintaining the plateau in the bulk region (illustrated by the purple line). The steady-state wavefunction profile [red circles in Fig.~\ref{Fig5}(e)], with its time- and space-resolved amplitude $\abs{\mathrm{Re}(\Phi_x)}$ shown in Fig.~\ref{Fig5}(d), closely matches the designed target profile of the TZM [blue lines in Fig.~\ref{Fig5}(e)].   A detailed discussion on the evolution into various designed wavefunction profiles is   provided  in Supplementary Note 7(A).   These results demonstrate  that, by leveraging  dynamical evolution, we can achieve any arbitrarily desired waveform across the entire lattice. 

\begin{figure*}[!tb]
	\centering
	\includegraphics[width=18cm]{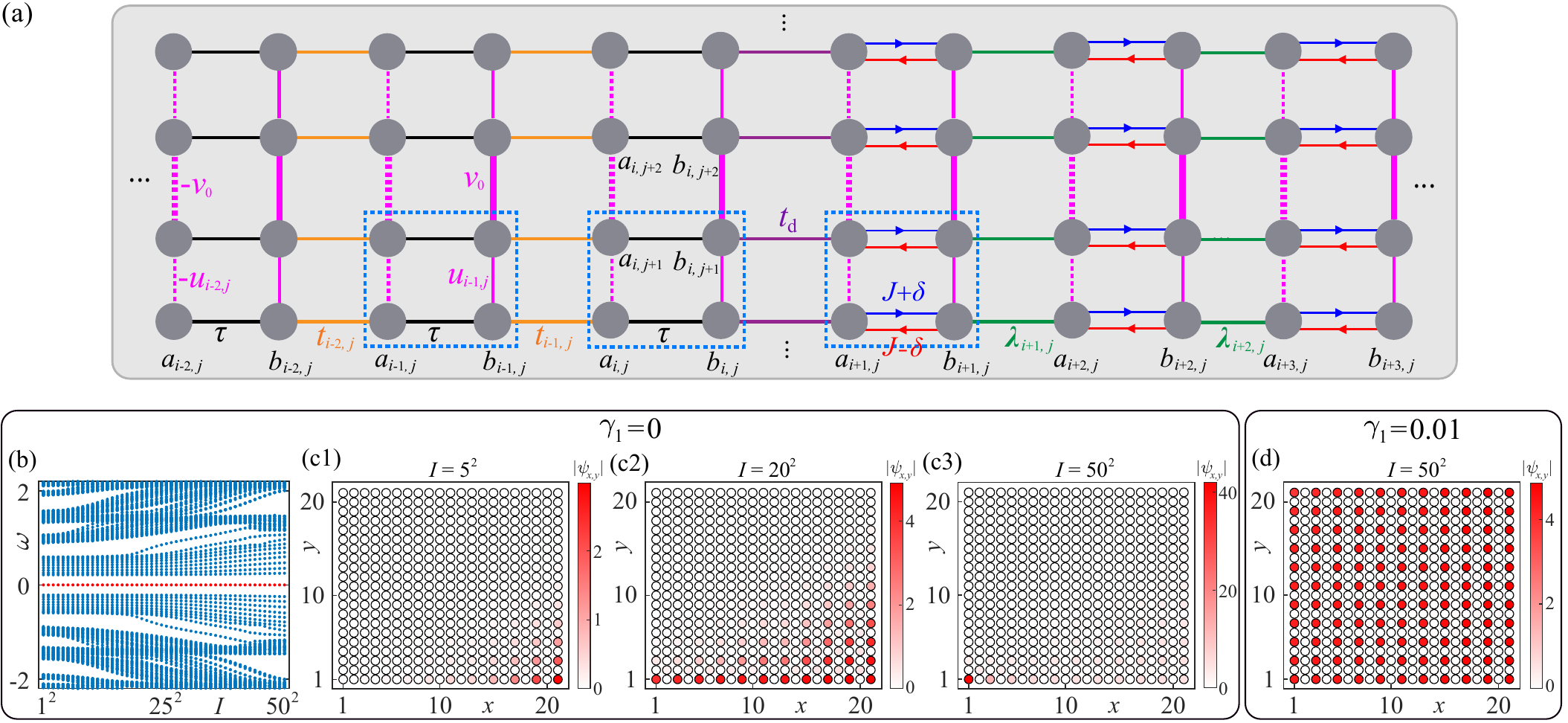}
	\caption{\textbf{Delocalization of topological zero modes (TZMs) in a 2D non-Hermitian nonlinear lattice}. (a) Schematic of a 2D non-Hermitian nonlinear interface model, formed by stacking 1D topological interface chains with staggered nonlinear hopping along the $y$ direction.  Each unit cell, indicated by a blue dashed box, contains four sublattices. The nonlinear intracell  hopping, along the $y$ direction, is given by $u_{i,j}=u_0 + \gamma_1 (\abs{\eta_{i,j}}^2 + \abs{\eta_{i,j+1}}^2)$  ($j \in \textrm{odd}$), where $\eta_{i,j} \in \{a_{i,j}, b_{i,j}\}$ are the state amplitudes of the two sites linked by the corresponding nonlinear coupling.  Dashed purple lines indicate hopping terms with negative signs along the $y$ direction. (b) Eigenfrequency $\omega$ versus $I =\sum_{i,j} (\abs{a_{i,j}}^2 + \abs{b_{i,j}}^2)$ for   $\gamma_1=0$,  where red dots mark TZMs. The corresponding spatial distributions $\abs{\psi_{x,y}}$ ($x$ and $y$ labels lattice site along the $x$ and $y$ directions)  of the TZMs for different $I$ are shown in (c1-c3). (d)  $\abs{\psi_{x,y}}$ for   $\gamma_1=0.01$, where the TZMs exhibit fully extended spatial distributions at $I=50^2$.  Other parameters used are  $J=1.5$, $\tilde{t}_{i,j}=\tilde{\lambda}_{i,j} = 1.5$, $\alpha=\beta=0.05$, $\tau=t_\text{d}=2.5$,  $\delta=1.2$, $u_0=0.2$,  $v_0=0.4$, $N=6$, and $L_x= L_y=21$.}\label{Fig6}
\end{figure*}

We now examine the stability of the evolved steady state under random noise. To incorporate the effects of noise, we introduce a disturbance $\ket{\Upsilon}$ to the evolved steady state $\ket{\Phi}$ at a certain evolution time. 
We then solve the dynamical evolution equation for the perturbed state $\ket{\phi} = \ket{\Phi} + \ket{\Upsilon}$ as
\begin{align}\label{SchrodingerEquationDynamicsdisorderSM}
	\frac{\partial \ket{\phi (t) }}{\partial t} = -i\left(\hat{\mathcal{H}} + \hat{\mathcal{H}}_\text{loss}\right)\ket{\phi (t)} + \xi \ket{\mathcal{P}} e^{-i\tilde{\omega} t},
\end{align}
where  the component of $\ket{\Upsilon}$ is randomly sampled in the range10 $[-3,~3]$. 

If the perturbed state $\ket{\phi}$ can return to the original  steady state $\ket{\Phi}$, it indicates that the evolved steady state is robust against random noise. To examine this return, we calculate the similarity function \cite{PhysRevLett.133.116602}, defined as 
\begin{align}\label{similarity}
	\chi(t) = \frac{ \abs{\braket{\Phi}{\phi(t)}}}{\sqrt{\braket{\Phi}{\Phi} \braket{\phi(t)}{\phi(t)} }},
\end{align}
where, when $\chi(t)$ reaches $1$, it indicates the stability of the evolved steady state under random noise.

We present the evolved similarity function $\chi(t)$ and the corresponding standard deviations $\sigma(t)$ for different pumping strength $\xi$, as shown in Fig.~\ref{Fig5}(f,g). The results are averaged over 200 independent realizations with a single-site pumping. As time evolves, $\chi(t)$ returns to $1$, and the corresponding standard deviation  $\sigma(t)$ approaches zero. This confirms the stable and fully-extended TZM achieved through external pumping.

Compared to the Hermitian case, the required pumping strength $\xi$ to achieve the desired long-range plateau is significantly lower in our non-Hermitian nonlinear interface model [see details in Supplementary Note 7(B) for further details on long-range pattern excitations for other wavefunction profiles]. In contrast, for the Hermitian interface model, the external pumping is insufficient to fully excite the predesigned pattern in a large-size lattice [see Fig.~\ref{Fig5}(h)]. This highlights how the interplay of nonlinearity, NHSE, and topology facilitates the dynamical preparation of long-range patterns with arbitrary wavefunction shapes. This capability could benefit future applications in reconfigurable topological photonic devices.

\vspace{.3cm}
\noindent \textbf{Delocalization of TZMs in a 2D non-Hermitian nonlinear lattice}

\noindent We now extend the 1D non-Hermitian nonlinear model to two dimensions (2D) by stacking 1D topological interface chains along the $y$ direction, as shown in Fig.~\ref{Fig6}(a). The blue dashed box marks a unit cell containing four sublattices, with odd and even indices $j$ distinguishing sites along $y$. Hopping amplitudes along the $y$ direction are both nonlinear and site-dependent, where dashed purple lines indicate hopping terms with negative signs. The nonlinear intracell coupling is given by
\begin{align}\label{2Dy1}
	u_{i,j} = u_0 + \gamma_1 (\abs{\eta_{i,j}}^2 + \abs{\eta_{i,j+1}}^2), ~j \in \textrm{odd},
\end{align}
where  $\eta_{i,j} \in \{a_{i,j}, b_{i,j}\}$ represents   the state amplitudes of the two sites linked by the corresponding nonlinear coupling, and the intercell coupling, characterized by strength $v_0$, is linear.  

Along the $x$ direction, for each chain, the intercell couplings, $t_{i,j}$ and $\lambda_{i,j}$, are nonlinear and given by
\begin{align}\label{2D1}
	t_{i,j} = \tilde{t}_{i,j} + \alpha (\abs{a_{i,j}}^2+\abs{b_{i-1,j}}^2),
\end{align}
\begin{align}\label{2D2}
	\lambda_{i,j} = \tilde{\lambda}_{i,j} + \beta (\abs{a_{i,j}}^2+\abs{b_{i-1,j}}^2).
\end{align}

In the absence of both nonlinearity and non-Hermiticity, the system reduces to an interface model based on the Benalcazar-Bernevig-Hughes (BBH) lattice \cite{Benalcazar2017}. The BBH model exhibits a higher-order topological phase characterized by the emergence of TZMs localized at the corners. In the linear Hermitian case, the corresponding 2D interface model supports TZMs localized at the ends of the interface.

When the nonlinearity along the $x$ direction is introduced with $\gamma_1 = 0$, the 2D non-Hermitian nonlinear interface model supports TZMs, as indicated by the red dots in Fig.~\ref{Fig6}(b). Owing to the strong NHSE, these TZMs are localized from the interface toward the right-bottom corner for weak nonlinearity at $I = 5^2$ [see Fig.~\ref{Fig6}(c1)]. As the nonlinear intensity increases to $I = 20^2$, the TZMs become more extended and predominantly occupy the bottom edge [see Fig.~\ref{Fig6}(c2)]. When the nonlinearity is further strengthened to $I = 50^2$, the modes do not spread across the entire lattice but instead become localized again [see Fig.~\ref{Fig6}(c3)]. In contrast, when nonlinearity is also introduced along the $y$ direction with $\gamma_1 = 0.01$, the TZMs extend over the entire 2D lattice at $I = 50^2$ [see Fig.~\ref{Fig6}(d)]. Notably, this extended phase emerges without the need to satisfy the constraint $\delta = \delta_\text{c} = \tilde{\lambda}_{i,j} - J$. These results demonstrate how the interplay between nonlinearity and non-Hermiticity enables flexible  control of topological mode localization in 2D lattices.

\vspace{.3cm}
\noindent {\large \textbf{Conclusions}}

\noindent In summary, we demonstrate how the intricate interplay of nonlinearity, NHSE and topology enables flexible control over the spatial profile of TMs. By coupling a Hermitian nonlinear SSH chain with a non-Hermitian one, we achieve full delocalization of the TM across the entire lattice. Unlike the delocalization induced by NHSE under critical conditions or the partial delocalization caused by nonlinearity in Hermitian systems, this triple synergy eliminates the need for precise parameter tuning to extend the TM across both chains. Furthermore, the TM waveform can be arbitrarily engineered by independently adjusting the nonlinearity strength in each chain. These extended non-Hermitian TMs exhibit robustness against disorder while maintaining dynamical stability and enabling long-range spatial patterns through external pumping. We also apply our approach to 2D systems, demonstrating its feasibility for delocalizing topological modes in higher-order topological phases. These results open potential avenues for applications in robust wave manipulation and nonlinear topological photonics, where tailored and disorder-resistant states are essential.

\vspace{.5cm}
\noindent {\large\textbf{Methods}}

\noindent \textbf{Spectral localizer in the nonlinear non-Hermitian  system}

\noindent Topological band theory establishes that topological invariants are global properties, intrinsically tied to the band structure and Bloch eigenstates of gaped systems. However, nonlinearity poses a significant challenge by introducing local effects that disrupt the spatial periodicity essential to traditional band theory.  To establish the topological origin of the zero modes in our nonlinear non-Hermitian system, we turn to the spectral localizer as a diagnostic tool \cite{PhysRevLett.131.213801,PhysRevB.108.195142,Cheng2023,PhysRevB.108.035107,PhysRevB.109.035425,Ochkan2024,Cerjan2023,LORING2015383}. 

The spectral localizer combines a system’s Hamiltonian and position operators through a non-trivial Clifford representation \cite{Cerjan2023,LORING2015383}, which is generally applied to Hermitian systems. To extend its application to our 1D non-Hermitian system, we employ a similarity transformation $\hat{S}$ that maps the  non-Hermitian  nonlinear Hamiltonian $\hat{\mathcal{H}}$ to an equivalent Hermitian Hamiltonian $\hat{\mathcal{H}}_\text{S}$. This transformation is defined as $\hat{\mathcal{H}}_\text{S} = \hat{S} \hat{\mathcal{H}} \hat{S}^{-1}$, ensuring compatibility with the spectral localizer framework. The matrix of the similarity transformation is given by
\begin{align}\label{SimilarityMatrixEqSM}
	\hat{S} = \left(\begin{matrix}
		\mathbf{I}_{2N}  &  \\
		& 0_{L-2N}
	\end{matrix}\right) + \left(\begin{matrix}
		0_{2N}  &  \\
		& \mathcal{R}_{L-2N}
	\end{matrix}\right),
\end{align}
where $\mathbf{I}$ is the identity matrix, and $\mathcal{R}$ is a diagonal matrix,  with dimensions $L-2N$, whose diagonal elements are $\{1,r,r,r^2,\cdots,r^{\frac{L-1}{2}-N-1},r^{\frac{L-1}{2}-N},r^{\frac{L-1}{2}-N}\}$ (where $L$ is odd). Here, $r=\sqrt{\abs{(J-\delta)/(J+\delta)}}$, and $\hat{\mathcal{H}}_\text{S}$ reads
\begin{align}\label{HamilSimilaritySM}
	\hat{\mathcal{H}}_\text{S} =&  \sum_{j\leq N} \left(\tau \ket{a_j} \bra{b_j} + t_{j-1} \ket{a_j} \bra{b_{j-1}} + \text{H.c.}\right)  \nonumber   \\
	&  + \sum_{j>N} \left[\sqrt{(J-\delta)(J+\delta)} \ket{a_j} \bra{b_j} +  \text{H.c.} \right] \nonumber   \\
	& + \sum_{j>N}  \left( \lambda_j \ket{a_{j+1}} \bra{b_j} + \text{H.c.} \right) \nonumber   \\
	&   + t_\text{d} \left(\ket{a_{N+1}} \bra{b_N} + \text{H.c.}\right).
\end{align}

The original nonlinear Schr\"{o}dinger equation, $\hat{\mathcal{H}} \ket{\psi} = \omega \ket{\psi}$,  now takes the form
\begin{align}\label{SchrodingerSimilarityEqSM}
	\hat{\mathcal{H}}_\text{S} \ket{\bar{\psi}} = \omega \ket{\bar{\psi}},
\end{align}
with $\ket{\bar{\psi}} = \hat{S} \ket{\psi}$. This allows us to leverage the information from the Hermitian Hamiltonian $\hat{\mathcal{H}}_\text{S}$ and its eigenmodes $\ket{\bar{\psi}}$ to characterize the topological properties of the non-Hermitian nonlinear system.

The nonlinear spectral localizer is a composite operator that incorporates the system's Hamiltonian $\hat{\mathcal{H}}_\text{S}$ accounting for its current occupations $\ket{\bar{\psi}}$ and position operators $\hat{X}$, using a nontrivial Clifford representation.  The 1D nonlinear spectral localizer is explicitly written as \cite{Cerjan2023,LORING2015383,PhysRevB.106.064109,Cheng2023,PhysRevLett.132.073803}
\begin{align}\label{localizerSM}
&	L_{\zeta\equiv(x,\bar{\omega})}(\hat{X},\hat{\mathcal{H}}_\text{S}) \nonumber   \\
	& = \left(\begin{matrix}
		0 & \eta(\hat{X}-x\mathbf{I}) - i (\hat{\mathcal{H}}_\text{S}-\bar{\omega}\mathbf{I}) \\
		\eta(\hat{X}-x\mathbf{I}) + i (\hat{\mathcal{H}}_\text{S}-\bar{\omega}\mathbf{I}) & 0
	\end{matrix}\right),
\end{align}
where the position matrix $\hat{X}$ is a diagonal matrix whose diagonal elements are $\{1,2,3,\cdots,L-1,L\}$, $\eta > 0$ is a scaling factor that ensures that the units of $\hat{X}$ and $\hat{\mathcal{H}}_\text{S}$ are compatible.  The pair $\zeta \equiv (x, \bar{\omega})$ serves as input for locally probing the topology in real space, with $x$ representing the spatial coordinate and $\bar{\omega}$ being the energy. This pair can be chosen either within or outside the system's spatial and spectral extent.

The spectral localizer  can be used to construct the relevant local topological invariant and to define the associated local gap. For a given $\zeta \equiv (x, \bar{\omega})$, the spectrum of the nonlinear spectral localizer, $\sigma(L_{\zeta}(\hat{X}, \hat{\mathcal{H}}_\text{S}))$, provides a measure of whether the system, when linearized around its current state $\ket{\bar{\psi}}$, supports a state $\ket{\bar{\Phi}}$ that is approximately localized near the point $(x, \bar{\omega})$, satisfying $\hat{\mathcal{H}}_\text{S} \ket{\bar{\Phi}} \approx \bar{\omega} \ket{\bar{\Phi}}$ and $\hat{X} \ket{\bar{\Phi}} \approx x \ket{\bar{\Phi}}$. In particular, if the smallest singular value of the spectral localizer,
\begin{align}\label{muMin}
	\mu_\zeta(\hat{X},\hat{\mathcal{H}}_\text{S}) = \text{min}\left[\abs{\sigma(L_{\zeta}(\hat{X},\hat{\mathcal{H}}_\text{S}))}\right],
\end{align}
is sufficient close to zero,  such an approximately localized state exists. Conversely, a large value indicates that the system does not support such a state. Thus, $\mu_\zeta$ can be heuristically understood as a local band gap \cite{Cheng2023}.

If the system respects chiral symmetry, the spectral localizer $L_{\zeta}(\hat{X},\hat{\mathcal{H}}_\text{S})$ can be further written in a reduced form as
\begin{align}\label{localizerReducedSM}
	\tilde{L}_{\zeta\equiv(x,\bar{\omega})}(\hat{X},\hat{\mathcal{H}}_\text{S}) = \eta(\hat{X}-x\mathbf{I})\hat{\Pi} + \hat{\mathcal{H}}_\text{S} - i \bar{\omega} \hat{\Pi},
\end{align}
where $\hat{\Pi}$ is the system's chiral operator, satisfying $\hat{\mathcal{H}}_\text{S} \hat{\Pi}= -\hat{\Pi} \hat{\mathcal{H}}_\text{S}$. Then, the local topological invariant $C_\zeta$ and local band gap $\mu_\zeta$ are given by \cite{Cheng2023}
\begin{align}\label{C1}
	C_\zeta(\hat{X},\hat{\mathcal{H}}_\text{S}) = \frac{1}{2} \text{Sig}(\tilde{L}_{\zeta}(\hat{X},\hat{\mathcal{H}}_\text{S})), 
\end{align}
\begin{align}\label{C2}
	\mu_\zeta(\hat{X},\hat{\mathcal{H}}_\text{S}) = \text{min}\left[\abs{\sigma(\tilde{L}_{\zeta}(\hat{X},\hat{\mathcal{H}}_\text{S}))}\right],
\end{align}
where $\text{Sig}$  is the signature of a matrix, i.e., its number of positive eigenvalues minus its number of negative ones, and $\sigma(\tilde{L}_{\zeta})$ is the eigenvalue of the reduced spectral localizer $\tilde{L}_{\zeta}$. For a system with an even or odd number of states, $C_\zeta$ is either integer or half-integer, and the changes in $C_\zeta$  are always integer-valued. Note that we consider $C_{(x,0)}(\hat{X},\hat{\mathcal{H}}_\text{S})$ for $\bar{\omega}=0$, reflecting the fact that the chiral symmetry can only protect states at the center of the system's eigenvalue spectrum.

The local band gap $\mu_\zeta$ and the local topological invariant $C_\zeta$ provide a consistent and complete description of a system's topology \cite{Cheng2023}. An intuitive physical interpretation of the spectral localizer's connection to a system's topology is that the invariant $C_\zeta(\hat{X}, \hat{\mathcal{H}}_\text{S})$ evaluates whether the matrices $\hat{\mathcal{H}}_\text{S}$ and $(\hat{X} - x\mathbf{I})$ can be smoothly deformed to the trivial atomic limit, where the Hamiltonian and position operator commute, without closing the spectral gap and violating a specific symmetry during the deformation. If $C_\zeta = 0$, this deformation is possible, indicating the system is topologically trivial at $\zeta = (x, \bar{\omega})$. However, if $C_\zeta \neq 0$, the deformation is obstructed, implying that the system is topologically nontrivial at $\zeta = (x, \bar{\omega})$. Furthermore, at the local site  $x_0$ for the specific eigenfrequency $\bar{\omega}$, when $C_{\zeta=(x_0, \bar{\omega})}$ changes value, the local band gap $\mu_{\zeta=(x_0, \bar{\omega})}$ necessarily closes with $\mu_{\zeta=(x_0, \bar{\omega})} = 0$. This gap closure corresponds to a state that is approximately localized near $x_0$ at the energy $\bar{\omega}$, thereby embodying the principle of bulk-boundary correspondence.

\vspace{.5cm}
\noindent{\large\textbf{Data availability} }

\noindent The numerical data presented in the figures is available at \url{https://zenodo.org/records/16812887}. Further rawdata is also available from the authors upon request.

\vspace{.5cm}
\noindent{\large\textbf{Code availability}} 

\noindent The codes are available upon reasonable request from the corresponding author.

\vspace{.5cm}
\noindent\textbf{\large References} 

\vspace{.5cm}

\noindent{\large \textbf{Acknowledgments}}

\noindent T.L. is grateful to Meng Xiao for valuable discussion. T.L. acknowledges the support from  the National Natural
Science Foundation of China (Grant No.~12274142),  the  Key Program of the National Natural Science Foundation of China (Grant No.~62434009), the Fundamental Research Funds for the Central Universities (Grant No.~2023ZYGXZR020), Introduced Innovative Team Project of Guangdong Pearl River Talents Program (Grant No.~2021ZT09Z109),  and the Startup Grant of South China University of Technology (Grant No.~20210012). Y.R.Z. thanks the support from the National Natural Science Foundation of China (Grant No.~12475017), Natural Science Foundation of Guangdong Province (Grant No.~2024A1515010398),  and the Startup Grant of South China University of Technology (Grant No.20240061). 
F.N. is supported in part by:  the Japan Science and Technology Agency (JST) [via the CREST Quantum Frontiers program Grant No. JPMJCR24I2, the Quantum Leap Flagship Program (Q-LEAP), and the Moonshot R$\&$D Grant No. JPMJMS2061].

\vspace{.5cm}

\noindent{\large \textbf{Author contributions}}  

\noindent T.L. and F.N. conceived the original concept and initiated the work.   Z.F.C., Y.C.W. and T.L. performed numerical calculations,  Y.R.Z. and T.L. analyzed and interpreted the results. All authors contributed to the  discussions of the results  and the development of the manuscript. T.L. and F.N. supervised the whole project.

\vspace{.5cm}
\noindent\textbf{\large Competing interests} 

\noindent The authors declare no competing interests.

\vspace{.5cm}
\noindent\textbf{\large Additional information} 

\noindent \textbf{Supplementary information.}  The online version contains supplementary material.

\clearpage \widetext
\begin{center}
	\section*{Supplemental Material for ``Versatile Control of Nonlinear Topological States in Non-Hermitian Systems"}
\end{center}
\setcounter{equation}{0} \setcounter{figure}{0}
\setcounter{table}{0} \setcounter{page}{1} \setcounter{secnumdepth}{3} \makeatletter
\renewcommand{\theequation}{S\arabic{equation}}
\renewcommand{\thefigure}{S\arabic{figure}}
\renewcommand{\bibnumfmt}[1]{[S#1]}
\renewcommand{\citenumfont}[1]{S#1}

\makeatletter
\def\@hangfrom@section#1#2#3{\@hangfrom{#1#2#3}}
\makeatother

\maketitle

\vspace{2mm}
\begin{center}\textbf{Supplementary Note 1: POSSIBLE EXPERIMENTAL SETUP}\end{center}
\vspace{2mm}
The non-Hermitian nonlinear lattices considered in this work can be feasibly implemented across a variety of experimental platforms, including photonic systems \cite{PhysRevA.100.063830SM,Sone2025SM} and electronic circuits \cite{Hadad2018SM,arXiv:2403.10590SM,arXiv:2505.09179SM}. Here, we focus on an electronic circuit platform that allows for both tunable nonreciprocal hopping \cite{Helbig2020SM,Guo2024SM,2025SM} and controllable nonlinearity \cite{Hadad2018SM,arXiv:2403.10590SM,arXiv:2505.09179SM}. In particular, the key ingredients of our model, i.e., nonreciprocal hopping and amplitude-dependent nonlinear hopping, can be effectively realized within this platform. In the following, we provide a detailed description of the circuit implementation of each hopping mechanism.

\begin{figure*}[!b]
	\centering
	\includegraphics[width=16cm]{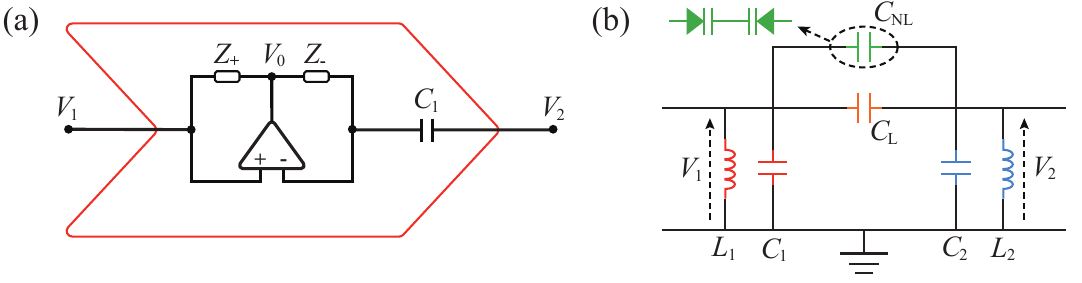}
	\caption{ (a) Experimental circuit realization of nonreciprocal hopping between two neighbor nodes via the negative impedance converters through current inversion (INIC), which consists of capacitor, resistor and operational amplifier.  (b) Experimental circuit realization of nonlinear hopping, which consists of two shut $LC$ resonators, a linear capacitor $C_\text{L}$ and a nonlinear capacitor $C_\text{NL}$. }\label{FigS1}
\end{figure*}

(i) The nonreciprocal hopping between neighboring nodes is implemented using negative impedance converter (INIC) circuits, which introduce direction-dependent current inversion \cite{Helbig2020SM, 2025SM}, as illustrated in Fig.~\ref{FigS1}(a). In this configuration, each pair of adjacent nodes is connected via a capacitor $C_\text{1}$ and an INIC. The INIC efficiently introduces an asymmetric capacitive coupling, exhibiting an equivalent capacitance of $\pm C_\text{2}$ depending on the direction of signal flow.

(ii) The nonlinear hopping between neighboring nodes is realized by utilizing a nonlinear capacitor\cite{Hadad2018SM}, as shown in Fig.~\ref{FigS1}(b). Specifically, we consider an interaction between two resonators coupled via a linear capacitor $C_\text{L}$ and a nonlinear capacitor $C_\text{NL}$. Following the derivation\cite{Hadad2018SM}, the dynamics of the system can be modeled using coupled-mode equations:
\begin{align}\label{model1}
	-j \frac{d a_\text{1}}{dt} = \omega_0 a_\text{1} + [\kappa+\nu(\abs{V_{C_\text{1}}-V_{C_\text{2}}})] a_\text{2},
\end{align}
\begin{align}\label{model2}
	-j \frac{d a_\text{2}}{dt} = \omega_0 a_\text{2} + [\kappa+\nu(\abs{V_{C_\text{1}}-V_{C_\text{2}}})] a_\text{1},
\end{align}
where $\omega_0$ is the resonance frequency, and the couplings consist of a linear term $\kappa=C_\text{L}/(C_\text{1}+C_\text{2})$ and a nonlinear term $\nu(V) = C_\text{NL}(\abs{V_{C_\text{1}}-V_{C_\text{2}}})/(C_\text{1}+C_\text{2})$, which depends on the voltage difference across the nonlinear capacitor. This structure allows the coupling strength to be modulated dynamically by the local voltage amplitude, thus realizing an effective nonlinear hopping mechanism.
Note that different forms of amplitude-dependent nonlinear coupling between two nodes have also been realized in circuit platforms \cite{arXiv:2403.10590SM,arXiv:2505.09179SM}. In addition, amplitude-dependent coupling can be implemented in optical systems using nonlinear fibers \cite{PhysRevA.100.063830SM,Sone2025SM}, where the coupling strength varies with the light intensity.

\vspace{2mm}
\begin{center}\textbf{Supplementary Note 2: IN-GAP NON-ZERO MODES}\end{center}
\vspace{2mm}

In the main text, we discuss the delocalization of in-gap topological zero modes (TZMs). Beyond these in-gap TZMs, the nonlinearity can also induce in-gap non-zero modes with eigenvalues that are not fixed at zero frequency. In this section, we provide details on the wavefunction distributions of these in-gap non-zero modes within the nonlinear eigenfrequency spectrum.

\begin{figure*}[!h]
	\centering
	\includegraphics[width=16cm]{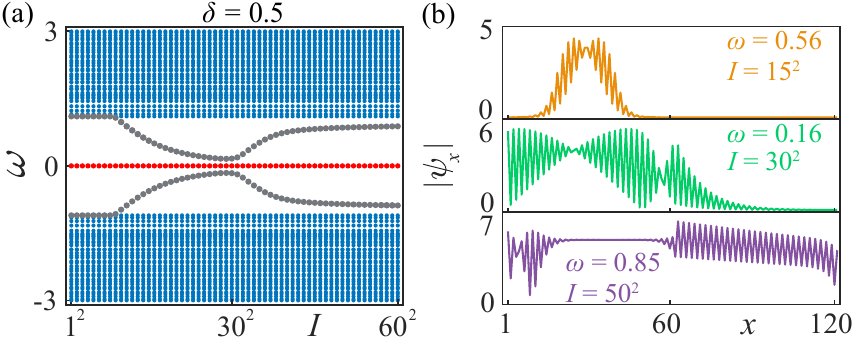}
	\caption{(a) Eigenfrequency spectrum $\omega$ versus $I=\sum_{j} (\abs{a_j}^2 + \abs{b_j}^2)$ for $\delta=0.5$, where the gray dots denotes the in-gap non-zero  modes and red dots mark the TZMs. The corresponding spatial distributions $\abs{\psi_x}$   of typical in-gap non-zero  modes for different $I$ are shown in (b). The parameters  used are   $J=1.5$, $\tau= \tilde{\lambda}_j=t_\text{d}=2.5$, $\tilde{t}_j = 1.0$,  $\alpha=0.05$, $\beta=0$,   $N=31$, and  $L=121$.  }\label{FigS2}
\end{figure*}

Figure \ref{FigS2}  shows the eigenfrequency spectrum $\omega$ versus  $I=\sum_{j} (\abs{a_j}^2 + \abs{b_j}^2)$, and the corresponding spatial distributions $\abs{\psi_x} $ of the in-gap non-zero modes [light black  dots in Fig.~\ref{FigS2}(a)] for  $\delta=0.5$ with $\alpha = 0.05$ and $\beta=0$. The in-gap non-zero modes emerge only when $I$ exceeds a certain value, and their  eigenvalues exhibit a strong dependence on $I$. Furthermore, as shown in Fig.~\ref{FigS2}(a), the in-gap non-zero modes originates from the  bulk modes. When $I$ is small, the in-gap non-zero modes are mainly localized at the Hermitian nonlinear chain. As $I$ increases, the in-gap non-zero modes become extended,  while the wavefunction profiles remain undefined or arbitrary in shape, as shown in Fig.~\ref{FigS2}(b). Due to the undefined waveform and unconstrained  eigenvalues of the in-gap non-zero modes for different $I$, in this work we focus exclusively  on the TZMs.

\begin{figure*}[!t]
	\centering
	\includegraphics[width=18cm]{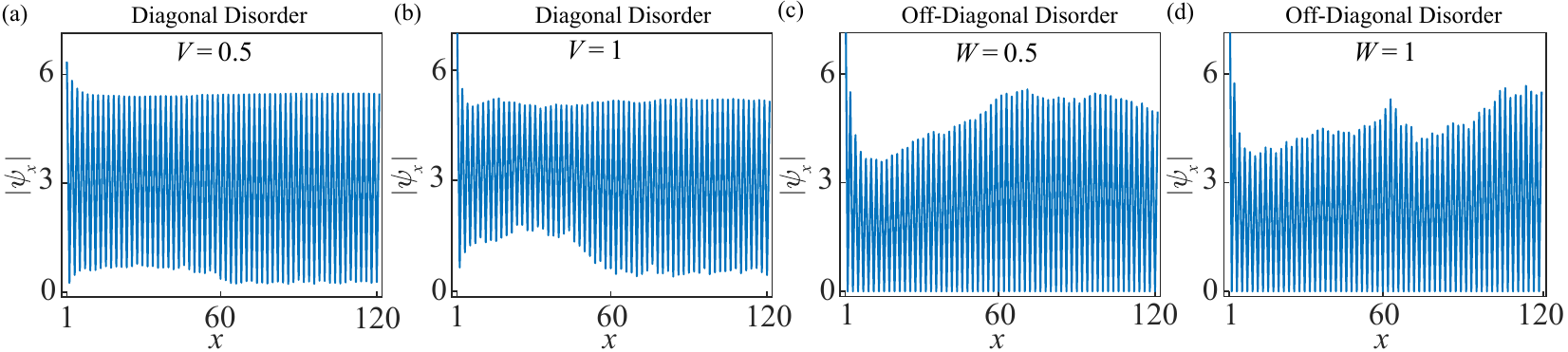}
	\caption{Spatial distributions $\abs{\psi_x}$ of the delocalized TZM  subject to a disordered onsite potential (a,b) and a disordered hopping (c,d) for  $\delta = \delta_\text{c} =1.0$,  $\alpha=0.05$ and $I=43^2$.  The  disordered onsite potential (i.e., diagonal disorder) and disordered hopping (i.e., off-diagonal disorder) are   randomly sampled within the respective ranges of $[-V/2,~V/2]$ and $[-W/2,~W/2]$, respectively. The other parameters used here are $J=1.5$,  $\beta=0$, $\tau=\tilde{\lambda}_j=t_\text{d}=2.5$,  $ \tilde{t}_j = 1.0$, $N=31$, and $L=121$. }\label{FigS3}
\end{figure*}

\vspace{2mm}
\begin{center}\textbf{Supplementary Note 3: EFFECTS OF DISORDER ON THE DELOCALIZED TZMS FOR $\delta=\delta_\text{c}$ AND $\beta=0$}\end{center}
\vspace{2mm}

As demonstrated in the main text, for $\delta=\delta_\text{c}$ and $\beta =0$, the TZM becomes delocalized, and can occupy both the entire Hermitian and non-Hermitian lattices due to the interplay of nonlinearity, the non-Hermitian skin effect (NHSE), and topology. Here, we discuss the robustness of the TZM wavefunction delocalization against disorder due to topological protection.

We investigate two types of random disorders, i.e., disordered on-site energy, $\hat{\mathcal{H}}_1$, and disordered hopping strength,  $\hat{\mathcal{H}}_2$,    each applied separately to the topological interface model.  The disordered Hamiltonian is written as $\hat{\mathcal{H}}_\textrm{dis} = \hat{\mathcal{H}} + \hat{\mathcal{H}}_j$ ($j=1,2$), with
\begin{align}\label{HamilSM}
	\hat{\mathcal{H}} =&  \sum_{j\leq N} \left(\tau \ket{a_j} \bra{b_j} + t_{j-1} \ket{a_j} \bra{b_{j-1}} + \text{H.c.}\right)   + \sum_{j>N} \left[(J-\delta) \ket{a_j} \bra{b_j} + (J+\delta) \ket{b_j} \bra{a_j} \right] \nonumber \notag \\
	& + \sum_{j>N}  \left( \lambda_j \ket{a_{j+1}} \bra{b_j} + \text{H.c.} \right)   + t_\text{d} \left(\ket{a_{N+1}} \bra{b_N} + \text{H.c.}\right),
\end{align}
\begin{align}\label{Hj1}
	\hat{\mathcal{H}}_1 = \sum_{j} \left(V_{a,j}\ket{a_j} \bra{a_j} + V_{b,j} \ket{b_j} \bra{b_j}\right),
\end{align}
\begin{align}\label{Hj2}
	\hat{\mathcal{H}}_2 = \sum_{j} \left(W_{1,j} \ket{a_j} \bra{b_j} + W_{2,j} \ket{a_{j+1}} \bra{b_j} + \text{H.c.}\right),
\end{align}
where the  disordered onsite potential $V_{a,j}$ and $V_{b,j}$ (i.e., diagonal disorder) and the disordered hopping $W_{1,j}$ and $W_{2,j}$ (i.e., off-diagonal disorder) are   randomly sampled within the respective ranges of $[-V/2,~V/2]$ and $[-W/2,~W/2]$, respectively.

Figure  \ref{FigS3} shows the spatial distributions $\abs{\psi_x}$ of the delocalized TZM  subject to the disordered onsite potential (a,b) and the disordered hopping (c,d) for  $\delta = \delta_\text{c} =1.0$,  $\alpha=0.05$ and $I=43^2$. The TZM remains extended even under strong disorder, demonstrating the robustness of topological protection in TZMs, even in the presence of strong nonlinear effects.

\begin{figure*}[!b]
	\centering
	\includegraphics[width=17.6cm]{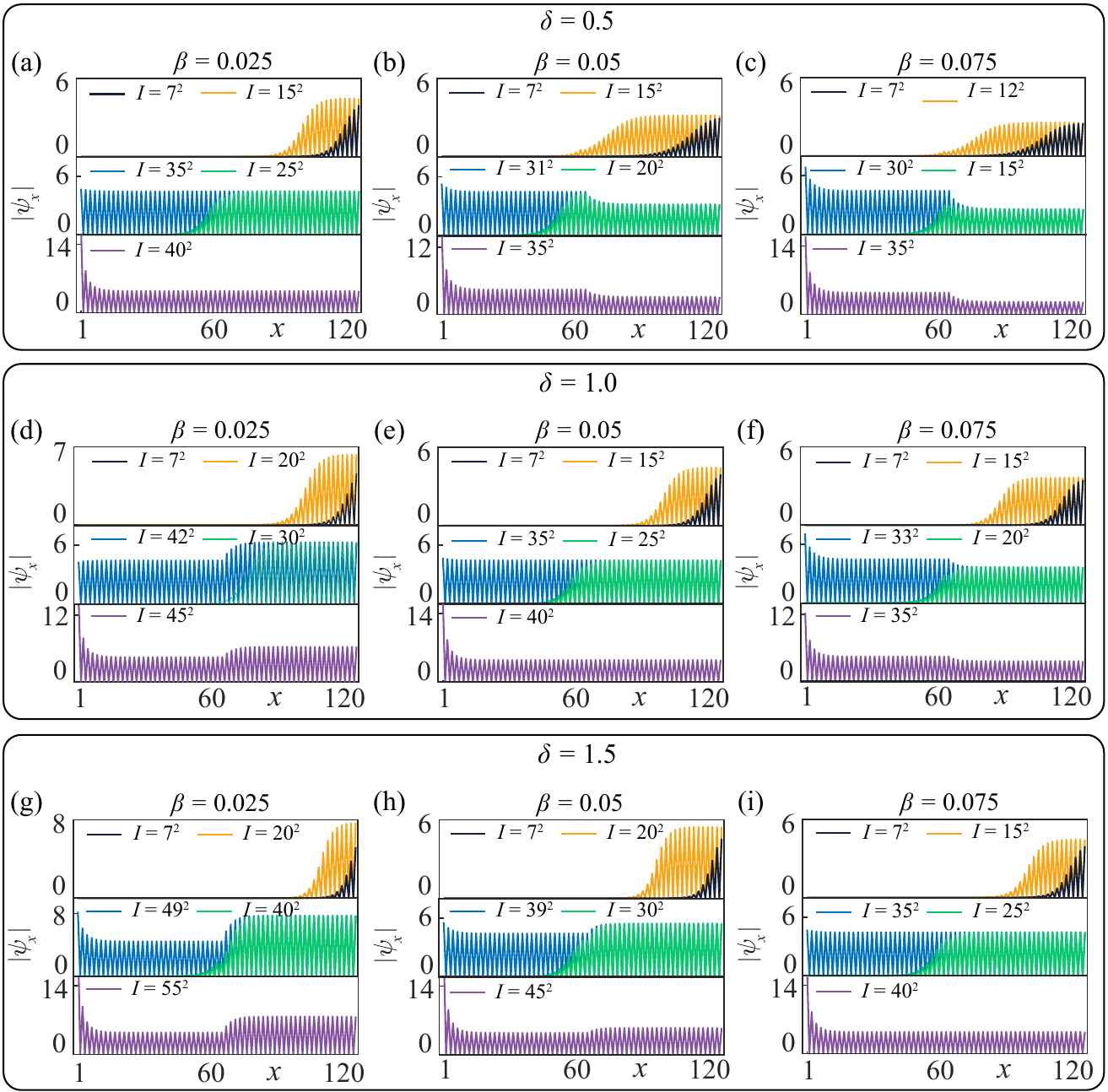}
	\caption{Spatial distributions $\abs{\psi_x}$ of the TZMs under various values of $I$ and the Kerr coefficient $\beta$ for (a-c) $\delta=0.5$, (d-f) $\delta=1.0$, and (g-i) $\delta=1.5$.   Other parameters used here are  $J=1.5$, $\alpha=0.05$, $\tilde{t}_j = \tilde{\lambda}_j = 1.5$, $\tau=t_\text{d}=2.5$, $N=31$, and $L=121$. }\label{FigS4}
\end{figure*}

\vspace{2mm}
\begin{center}\textbf{Supplementary Note 4: WAVEFUNCTION PROFILES OF TZMS FOR DIFFERENT KERR NONLINEAR COEFFICIENTS $\beta$ AND NONRECIPROCAL HOPPING AMPLITUDE $\delta$}\end{center}
\vspace{2mm}

This section provides a detailed exploration of the nonlinearity-driven control of TZMs. We begin by now presenting the nonlinear eigenequation for a TZM with zero eigenfrequency, $\omega=0$, expressed as 
\begin{align}\label{NonlinearEigenequation1}
	\tau a_j + \left(\tilde{t}_j + \alpha a^2_{j+1} \right) a_{j+1} = 0,~~~~ j<N,
\end{align}
\begin{align}\label{NonlinearEigenequation2}
	\tau a_N + t_\text{d} a_{N+1} = 0,~~~~ j=N,
\end{align}
\begin{align}\label{NonlinearEigenequation3}
	(J+\delta) a_j + \left(\tilde{\lambda}_j + \beta a^2_{j+1} \right) a_{j+1} = 0,~~~~ j>N,
\end{align}
and   $b_j = 0$. We now focus on parameters that deviate from the linear critical condition  $\delta_\text{c}=\lambda-J$ below, and begin by examining the case where $\tilde{t}_j = t $ and $\tilde{\lambda}_j = \lambda$ are constants. To this end, we consider the case where the Hermitian chain is in the topological trivial regime with $\tau > t$, and the non-Hermitian chain in the topological nontrivial regime, with  $J \in [-\sqrt{\delta^2 + \lambda^2},~\sqrt{\delta^2 + \lambda^2}]$ for $\abs{\lambda} \geq \abs{\delta}$, when $\beta = 0$ \cite{ShunyuYao2018SM}. Without loss of generality, we further assume $  (J + \delta) > \lambda $.

Figure \ref{FigS4} plots spatial distributions $\abs{\psi_x}$ of the TZMs under various values   of $I$, $\beta$ and $\delta$.  According to Eqs.~(\ref{NonlinearEigenequation1})-(\ref{NonlinearEigenequation3}), we can infer that the TZM is exponentially localized at the right boundary due to the NHSE when the nonlinear intensity $I$ is small, as indicated by the black lines in Fig.~\ref{FigS4}. As $I$ increases, the nonlinear term   $\tilde{\lambda}_j + \beta a^2_{j+1}$ in Eq.~(\ref{NonlinearEigenequation3}) grows accordingly. This behavior, when analyzed through Eqs.~(\ref{NonlinearEigenequation1})–(\ref{NonlinearEigenequation3}), suggests that the wavefunction of the TZM gradually extends from the right boundary into the bulk, progressively occupying more bulk sites. This transition is illustrated by the numerical results shown by the orange and green lines in Fig.~\ref{FigS4}. Eventually, under strong nonlinear intensity $I$, the TZM extends to occupy the entire lattice of the interface model, forming either one or two plateaus depending on the values of $\delta$ and $\beta$, as illustrated by the blue lines in Fig.~\ref{FigS4}. When $I$ is further increased beyond this threshold value, the wavefunctions become concentrated at the left end of the chain, while the plateaus are maintained (see purple lines in Fig.~\ref{FigS4}). Moreover, using  Eqs.~(\ref{NonlinearEigenequation1})-(\ref{NonlinearEigenequation3}), we can infer that the spatial distribution $\abs{\psi_x}$ of the TZM should remain flat across the lattice, except near the interface, because any significant deviation from this flatness would cause the wavefunction to diverge in the thermodynamic limit.

\begin{figure*}[!b]
	\centering
	\includegraphics[width=18cm]{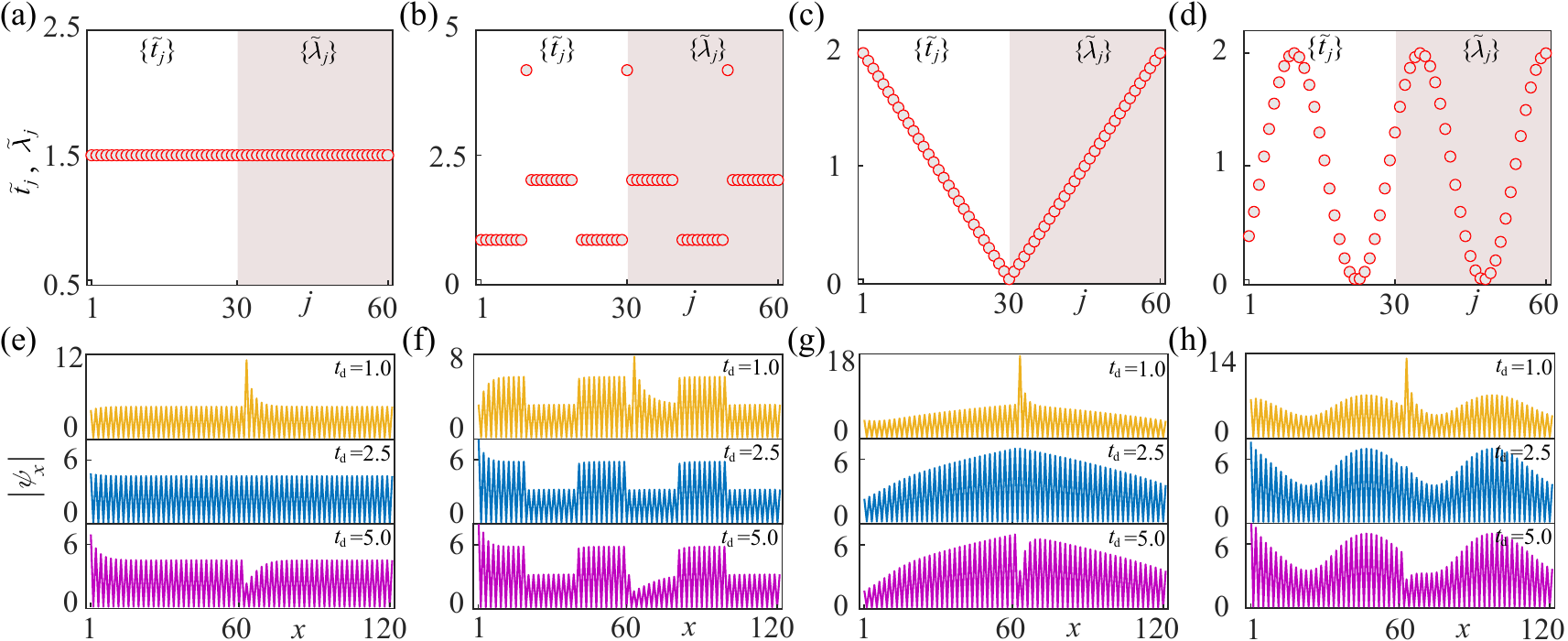}
	\caption{(a-d) Distributions of the hopping energies $\tilde{t}_j$ and $\tilde{\lambda}_j$ used to create flat, square, isosceles triangle, and cosine-shaped wavefunction profiles, respectively. Unfilled and filled regions correspond to the Hermitian and non-Hermitian regimes of the chains. The spatial distributions $\abs{\psi_x}$ of the  TZMs  are shown in (e-h), where the extended wavefunctions exhibit (e) a flat profile, (f) a square profile, (g) an isosceles triangle profile, and (h) a cosine profile. The state amplitude of the TZM around the interface sites can be tuned by adjusting the inter-chain coupling strength $t_\text{d}$ [see each row in (e-h)]. Other parameters used here are  $J=1.5$, $\delta=1.0$, $\alpha=\beta=0.05$, $\tau=2.5$, $N=31$ and $L=121$.}\label{FigS5}
\end{figure*}

Based on the above discussion, in the bulk regime under strong nonlinearity, the spatial distribution of the TZM tends to become uniform along each chain. Therefore, by applying Eqs.~(\ref{NonlinearEigenequation1})–(\ref{NonlinearEigenequation3}), the plateau heights in the bulk regimes of the non-Hermitian and Hermitian chains can be derived as follows: for $j>N$, the state amplitude of the TZM is $\abs{a_\text{R}} = \sqrt{(J+\delta-\lambda)/\beta}$, and for $j<N$, the state amplitude of the TZM is $\abs{a_\text{L}} = \sqrt{(\tau-t)/\alpha}$, respectively. By varying  $\beta$ for different values of $\delta$, the relative  plateau heights between the Hermitian and non-Hermitian chains can be tuned flexibly (see Fig.~\ref{FigS4}). Specifically, by choosing  $\delta$ and $\beta$ such that  $\abs{a_\text{L}} = \abs{a_\text{R}}$, a uniform plateau across the entire lattice can be achieved.

\vspace{2mm}
\begin{center}\textbf{Supplementary Note 5: DETAILS ON ARBITRARY DESIGN OF WAVEFUNCTION PROFILES OF TZMS}\end{center}
\vspace{2mm}

For constant values of $\tilde{t}_j$ and $\tilde{\lambda}_j$, we can obtain a  flat wavefunction profile of the TZM across the entire lattice. As demonstrated in the main text, the wavefunction profiles across the lattice can be designed arbitrarily. In this section, we provide detailed descriptions of the distributions of $\tilde{t}_j$ and $\tilde{\lambda}_j$ for wavefunction profiles with square, isosceles triangle, and cosine shapes.   Additionally, we examine the impact of disorder on these distributions.

\vspace{2mm}
\begin{center}\textbf{A. Arbitrary  wavefunction profile by engineering the hopping energies $\tilde{t}_j$ and $\tilde{\lambda}_j$}\end{center}
\vspace{2mm}

The wavefunction profiles of the extended TZMs across the entire lattice can be freely designed by engineering   distributions of ${\tilde{t}_j}$ and ${\tilde{\lambda}_j}$. Figure \ref{FigS5}(a-d) illustrates the distributions of ${\tilde{t}_j}$ and ${\tilde{\lambda}_j}$ employed to generate flat, square, isosceles triangle, and cosine-shaped wavefunction profiles \cite{PhysRevLett.133.116602SM}, respectively. The spatial distributions $\abs{\psi_x}$ of the  TZMs  are shown in \ref{FigS5}(e-h), where the extended wavefunctions exhibit (e) a flat profile, (f) a square profile, (g) an isosceles triangle profile, and (h) a cosine profile. The state amplitude of the TZM around the interface sites can be tuned by adjusting the inter-chain coupling strength $t_\text{d}$ [see each row in (e-h)]. Specifically, at the interface site,   we have $\abs{a_{N+1}} = \tau \abs{a_N}/t_d$, which satisfies $\abs{a_{N+1}} > \abs{a_N}$ for $t_\text{d}<\tau$ and $\abs{a_{N+1}} < \abs{a_N}$ for $t_\text{d}>\tau$, as manifested in Fig.~\ref{FigS5}(e-h).

\begin{figure*}[!b]
	\centering
	\includegraphics[width=18cm]{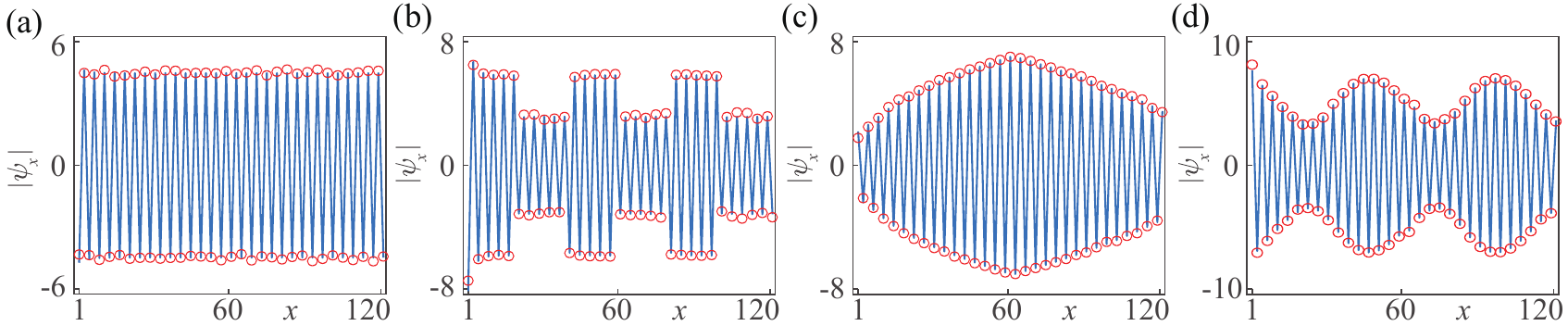}
	\caption{Spatial distributions $\abs{\psi_x}$ of the TZMs  for wavefunction profiles with (a) flat, (b) square, (c) isosceles triangle, and (d) cosine shapes, subject to random disorder applied to the hopping energies $\tilde{t}_j \to  \tilde{t}_j (1+ W_j)$ and $\tilde{\lambda}_j \to  \tilde{\lambda}_j (1+ V_j)$. The blue lines indicate the state distributions without perturbation, while the red circles depict the distributions in the presence of disorder. Other parameters used are   $J=1.5$, $\delta=1.0$, $W=0.2$, $\alpha=\beta=0.05$, $\tau=t_\text{d}=2.5$, $N=31$ and $L=121$. }\label{FigS6}
\end{figure*}

\vspace{2mm}
\vspace{2mm}
\begin{center}\textbf{B. Effect of disorder on arbitrarily-designed wavefunction profiles}\end{center}
\vspace{2mm}

To demonstrate the robustness of arbitrarily-designed wavefunction profiles for the TZMs across the entire lattice under the influence of disorder, we introduce random perturbations to the hopping  energies. The hopping energies are modified as $\tilde{t}_j \to \tilde{t}_j (1 + W_j)$, and  $\tilde{\lambda}_j \to \tilde{\lambda}_j (1 + V_j)$, where $W_j$ and $V_j$ are independent random variables uniformly distributed over the range $[-W/2,~W/2]$. Here, $W$ quantifies the disorder strength, providing a controlled parameter to evaluate the stability and resilience of the wavefunction profiles against spatially distributed random disorder. 

Figure \ref{FigS6} illustrates the spatial distributions $\abs{\psi_x}$ of the TZMs for wavefunction profiles with shapes (a) flat, (b) square, (c) isosceles triangle, and (d) cosine, both in the absence of disorder (blue lines) and in the presence of disorder (red circles). The disordered state distributions show negligible deviations from the unperturbed ones, highlighting the remarkable robustness of the designed TZMs against  disorder due to their topological nature.

\begin{figure*}[!h]
	\centering
	\includegraphics[width=18cm]{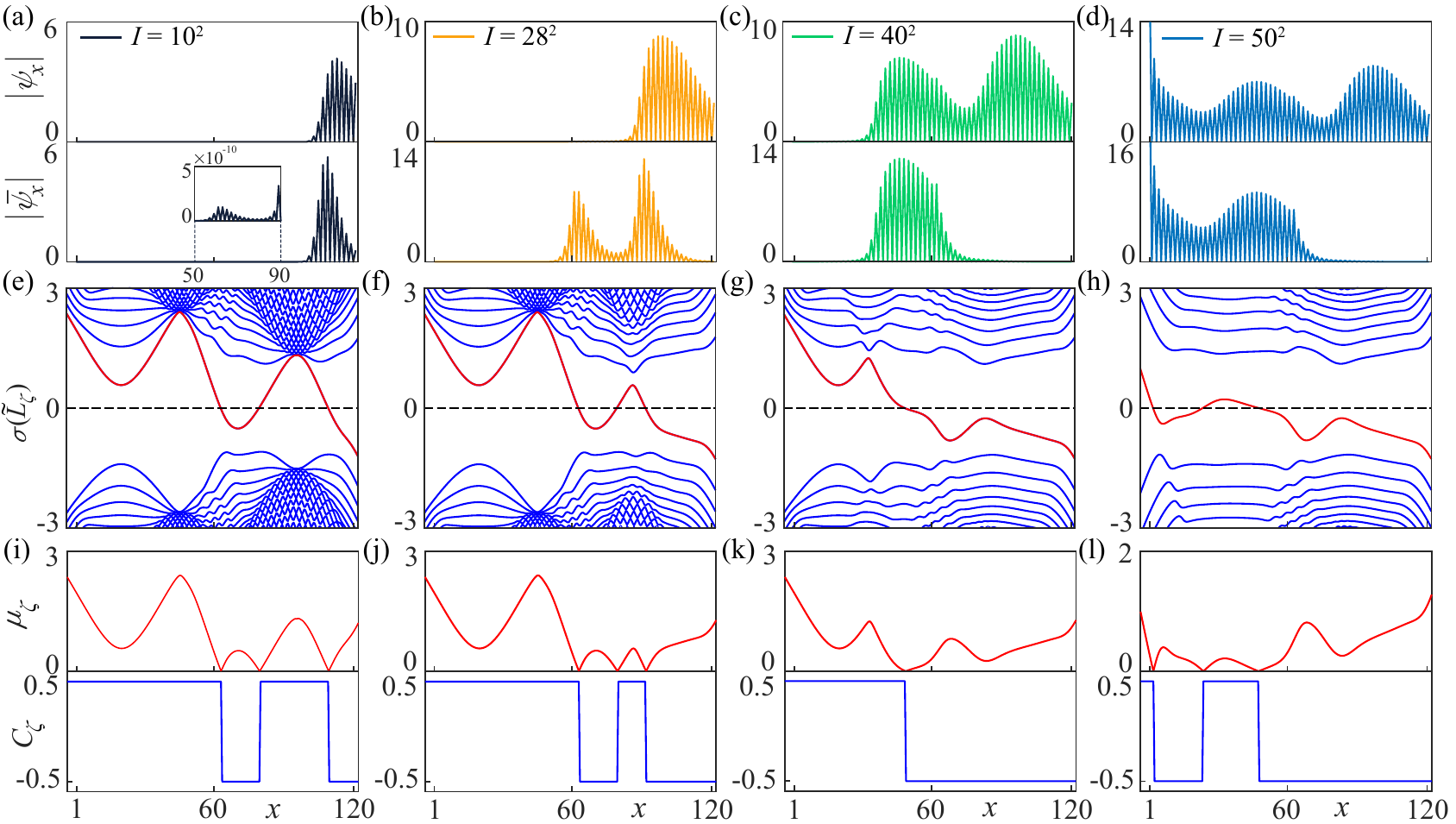}
	\caption{(a-d) Spatial distribution  $\abs{\psi_x}$ of the TZM   (upper panel) and their transformed counterparts $\abs{\bar{\psi}_x} = \abs{\hat{S} \psi_x}$ under a similarity transformation (lower panel) for different $I$. (e-h) Site-resolved eigenvalues $\sigma(\tilde{L}_\zeta)$ of the reduced spectral localizer $\tilde{L}_\zeta$ using the similarity-transformed Hamiltonian $\hat{\mathcal{H}}_\text{S}$. (i-l) Site-resolved local gap $\mu_\zeta$ (upper panels) and topological invariant $C_\zeta$ (lower panels)   for different $I$. The distributions of $\tilde{t}_j$ and $\tilde{\lambda}_j$, corresponding to the cosine-shaped TZM, are shown in Fig.~\ref{FigS5}(d). The pair $\zeta \equiv (x, \bar{\omega})$ is set as $\bar{\omega}=0$, and $\eta=0.2$. The other parameters used are $J=1.5$, $\delta=0.5$, $\alpha=0.05$, $\beta=0.025$, $\tau=t_\text{d}=2.5$, $N=31$ and $L=121$.  }\label{FigS7}
\end{figure*}

\begin{figure*}[!h]
	\centering
	\includegraphics[width=16cm]{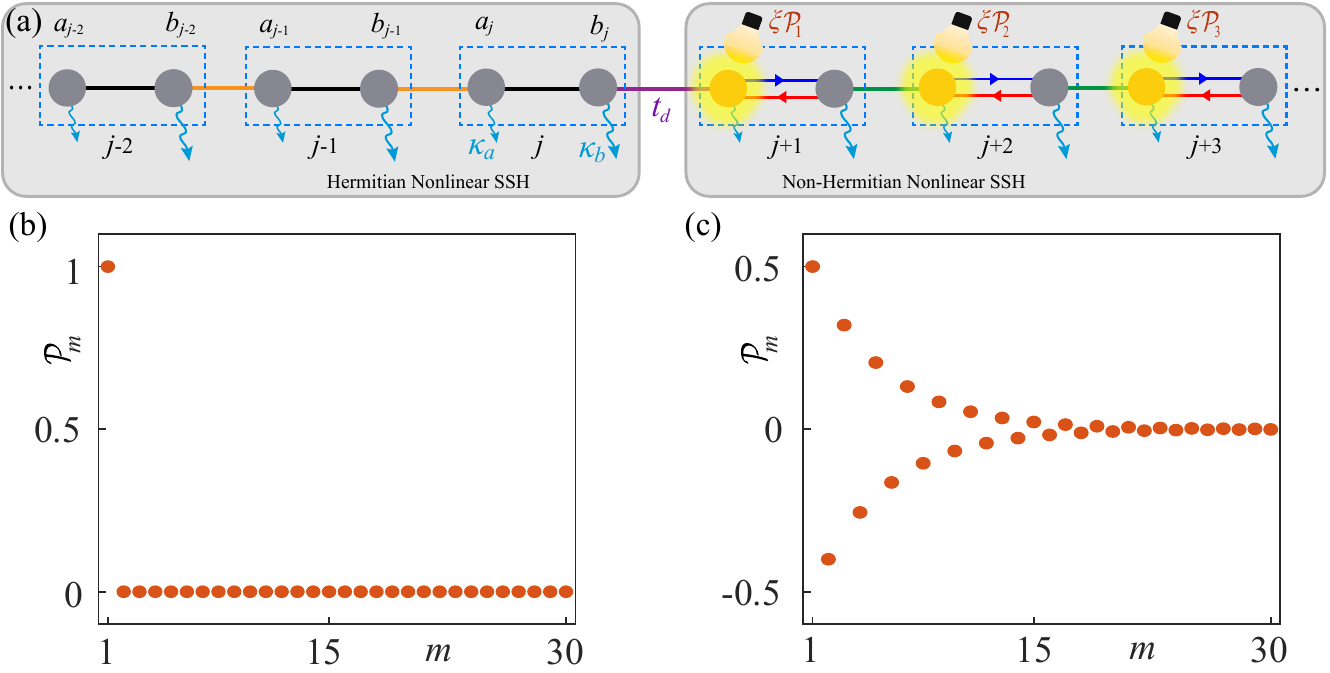}
	\caption{(a) Schematic showing external pumping applied to the $a$-sites of the non-Hermitian chain, highlighted by yellow spots. The intensity distributions of pumping sources are represented by $\xi \mathcal{P}_m$, with $\xi$ being  pumping strength. The cyan wave arrows indicate onsite losses in the two sublattices, labeled  $\kappa_\text{a}$ and $\kappa_\text{b}$.   (b, c)  Distributions of two types of pumping sources $\ket{\mathcal{P}_m}$. }\label{FigS8}
\end{figure*}

\vspace{2mm}
\begin{center}\textbf{Supplementary Note 6: SPECTRAL LOCALIZER IN THE NONLINEAR NON-HERMITIAN SYSTEM}\end{center}
\vspace{2mm}

In the main text, we present the  local band gap $\mu_\zeta$ and the local topological invariant $C_\zeta$ for the flat-shaped TZMs. Here, we discuss the topological properties of  the cosine-shaped TZMs.

Figure \ref{FigS7}(a-d) shows
the spatial distribution  $\abs{\psi_x}$ of the TZM   (upper panel) and their transformed counterparts $\abs{\bar{\psi}_x} = \abs{\hat{S} \ket{\psi}}$ under a similarity transformation (lower panel) for different $I$. The extended cosine-shaped TZM across the entire lattice is observed as the nonlinear intensity  $I$ increases. We then calculate the site-resolved eigenvalues $\sigma(\tilde{L}_{\zeta})$ of the reduced spectral localizer $\tilde{L}_\zeta$ for $\bar{\omega}=0$ using the similarity-transformed Hamiltonian $\hat{\mathcal{H}}_\text{S}$, as shown in Fig.~\ref{FigS7}(e-h). The spectrum $\sigma(\tilde{L}_{\zeta})$ crosses the zero energy line (black dotted line), as indicated by the red curves. The site-resolved local gap $\mu_\zeta$ (upper panels) and topological invariant $C_\zeta$ (lower panels) are shown in Fig.~\ref{FigS7}(i-l). For each nonlinear intensity $I$, as $x$ varies, the local band gap $\mu_\zeta$ closes, accompanied by a change in the local topological invariant $C_\zeta$. This indicates the presence of a TZM localized near the corresponding $x$, reflecting the  bulk-boundary correspondence. Furthermore, the regions where the local band gap $\mu_\zeta$ is closest to zero differ for each intensity, signifying the emergence of extended TZMs.

\vspace{2mm}
\begin{center}\textbf{Supplementary Note 7: DETAILS OF EXTENDED TZMS UNDER EXTERNAL PUMPING}\end{center}
\vspace{2mm}

\begin{center}\textbf{A. Dynamical evolution under external pumping}\end{center}
\vspace{2mm}

\begin{figure*}[!t]
	\centering
	\includegraphics[width=17.6cm]{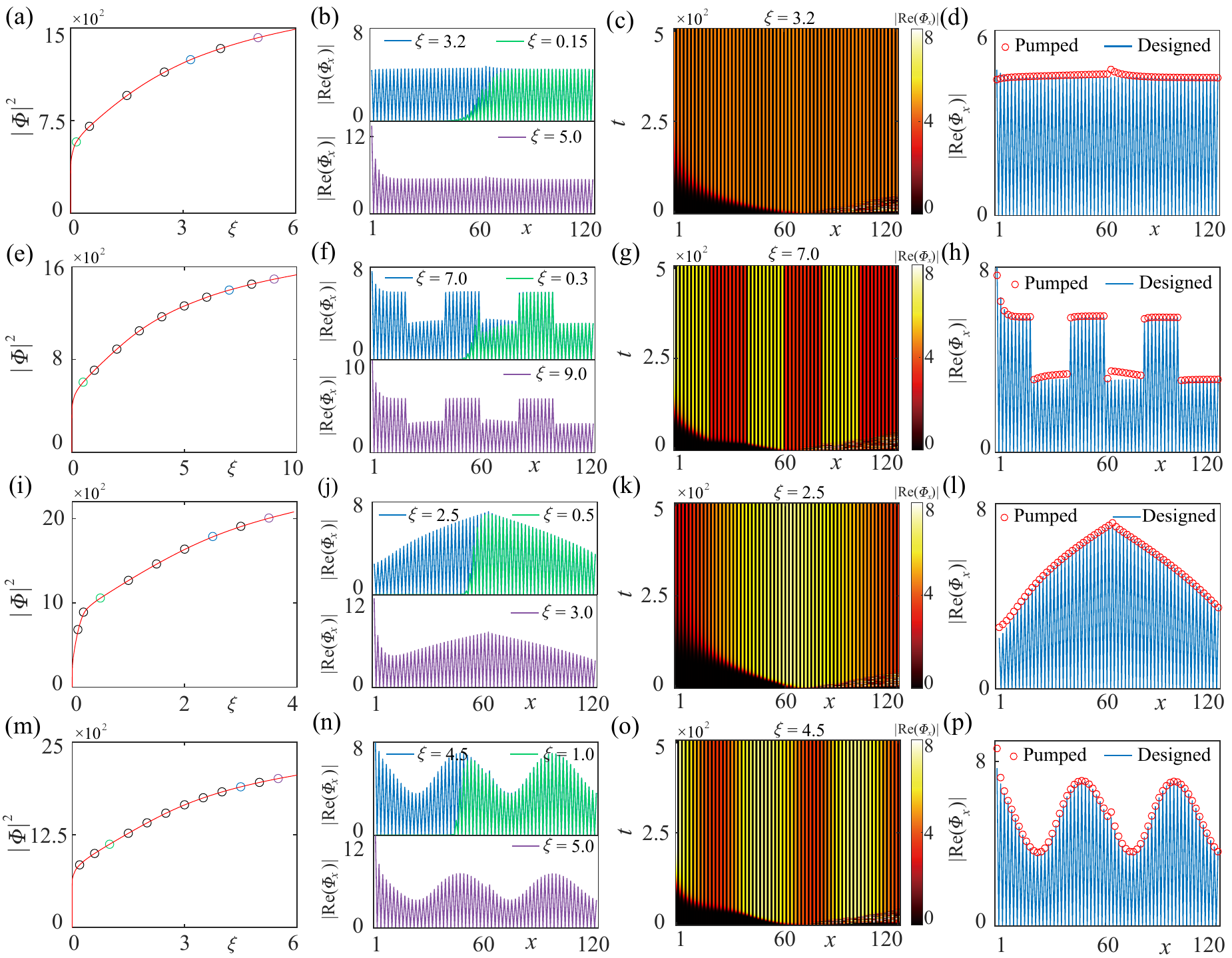}
	\caption{(a,e,i,m) Intensity $\abs{\Phi}^2$ of the evolved steady state versus $\xi$ for different designed wavefunction profiles with  flat (a),  square (e),  isosceles triangle (i) and  cosine shapes (m). The circles represent results from the evolution equation Eq.~(\ref{SchrodingerEquationDynamicsSM}), which closely match the results (red curve) obtained from the self-consistent nonlinear equation Eq.~(\ref{steadystateSM}). (b,f,j,n) Wavefunction profile $\abs{\text{Re}(\Phi_x)}$ of the  evolved steady state  versus $\xi$ for different designed wavefunction profiles. (c,g,k,o) Time- and space-resolved $\abs{\text{Re}(\Phi_x)}$  for different $\xi$ [the value of $\xi$ marked by the blue circle in (a,e,i,m)]. The corresponding steady-state result is  highlighted by red circles  in (d,h,l,p).  In (d,h,l,p), the steady-state wavefunction profile (red circles) closely matches the designed one (blue line) of the TZM.  The distribution of $\mathcal{P}_m$ is shown in Fig.~\ref{FigS8} (c). The parameters used are  $J=1.5$, $\delta=1.0$, $\alpha=\beta=0.05$, $\tau=t_\text{d}=2.5$, $\kappa_\text{a}=0.01$, $\kappa_\text{b}=0.5$, $\tilde{\omega}=0$, $N=31$ and $L=121$. }\label{FigS9}
\end{figure*}


This section provides details on the preparation of TZMs via external pumping. As shown in the main text, nonlinear models can exhibit distinctive dynamical properties that depend on how intensity levels are reached, enabling intrinsic control on TZMs through external pumping. Using the external pumping scheme illustrated in Fig.~\ref{FigS8}(a),  we solve the dynamical evolution equation, shown in the main text, with
\begin{align}\label{SchrodingerEquationDynamicsSM}
	\frac{\partial \ket{\varphi}}{\partial t} = -i\left(\hat{\mathcal{H}} + \hat{\mathcal{H}}_\text{loss}\right)\ket{\varphi} + \xi \ket{\mathcal{P}} e^{-i\tilde{\omega} t},
\end{align}
where $\hat{\mathcal{H}}_\text{loss} = \sum_{j} (-i \kappa_\text{a} \ket{a_j} \bra{a_j} - i \kappa_\text{b} \ket{b_j}  \bra{b_j})$, denotes onsite losses in the two sublattices, which contributes to stabilizing the excitation. The pumping sources $\ket{\mathcal{P}} \equiv (0,\cdots,\mathcal{P}_m,\cdots)^T$ are only applied to the $a$-sites of the non-Hermitian chain [see Fig.~\ref{FigS8}(a)], with the pumping frequency denoted by $\tilde{\omega}$, and the pumping strength $\xi$. 

Exciting the system initially in the vacuum state at the pumping frequency $\tilde{\omega}$, the evolved steady state $\ket{\Phi}$ from Eq.~(\ref{SchrodingerEquationDynamicsSM}) satisfies
\begin{align}\label{steadystateSM}
	(\hat{\mathcal{H}} + \hat{\mathcal{H}}_\text{loss} - \tilde{\omega}) \ket{\Phi} = -i \xi \ket{\mathcal{P}}.
\end{align}
Given $\tilde{\omega} = 0$ and $\xi\ket{\mathcal{P}}$, the steady state $\ket{\Phi}$ for the excited TZMs can be numerically obtained using a self-consistent method applied to Eq.~(\ref{steadystateSM}). We consider two distinct distributions of $\mathcal{P}_m$ \cite{PhysRevLett.133.116602SM}, as shown in Figs.~\ref{FigS8}(b) and (c).

In the main text, we discussed the excitation of the TZM with a flat wavefunction profile through the external pumping, specifically using a single-site pumping scheme with  $\{\mathcal{P}_m\}=\delta_{m,1}$,  as shown in Fig.~\ref{FigS8}(b). Here, we extend our discussion to the realization of other wavefunction profiles by employing a more generalized pumping scheme, as depicted in Fig.~\ref{FigS8}(c). This approach allows for enhanced control over the spatial structure of the wavefunction \cite{PhysRevLett.133.116602SM}.

Figure \ref{FigS9}(a,e,i,m) presents the intensity distribution $\abs{\Phi}^2$ of the evolved steady state at the resonant pumping frequency $\tilde{\omega} = 0$, plotted as a function of the pumping strength $\xi$. The results are shown for different target wavefunction profiles: flat (a), square (e), isosceles triangle (i), and cosine (m). The numerical results obtained from the dynamical evolution of Eq.~(\ref{SchrodingerEquationDynamicsSM}) (black circles) closely match the steady-state solutions (red curves) derived from the self-consistent nonlinear equation Eq.~(\ref{steadystateSM}). The corresponding spatial profiles of the evolved steady-state wavefunction $\abs{\text{Re}(\Phi_x)}$ at representative pumping strengths $\xi$ are illustrated in Fig.~\ref{FigS9}(b,f,j,n). As $\xi$ increases, the wavefunction progressively expands across the entire lattice while retaining the desired spatial profile. Initially, the excitation is localized near the right boundary (not shown here) due to the NHSE, but with increasing pumping strength, the steady-state profile extends leftward (green line), approaching the designed target shape (blue line). Even under strong pumping, the system remains capable of stabilizing into the predefined wavefunction form (purple line) in spite of a more density accumulation at the left boundary.

To further examine the dynamical evolution toward the steady state, figure \ref{FigS9}(c,g,k,o) displays the time- and space-resolved evolution of $\abs{\text{Re}(\Phi_x)}$. The steady-state wavefunction profile (red circles) is explicitly highlighted in Fig.~\ref{FigS9}(d,h,l,p), demonstrating its excellent agreement with the intended target profile (blue line). These results confirm that the external pumping effectively drives the system into the designed wavefunction profiles across the entire lattice due to the triple interplay of the nonlinearity, topology and NHSE.

\begin{figure*}[!tb]
	\centering
	\includegraphics[width=17.6cm]{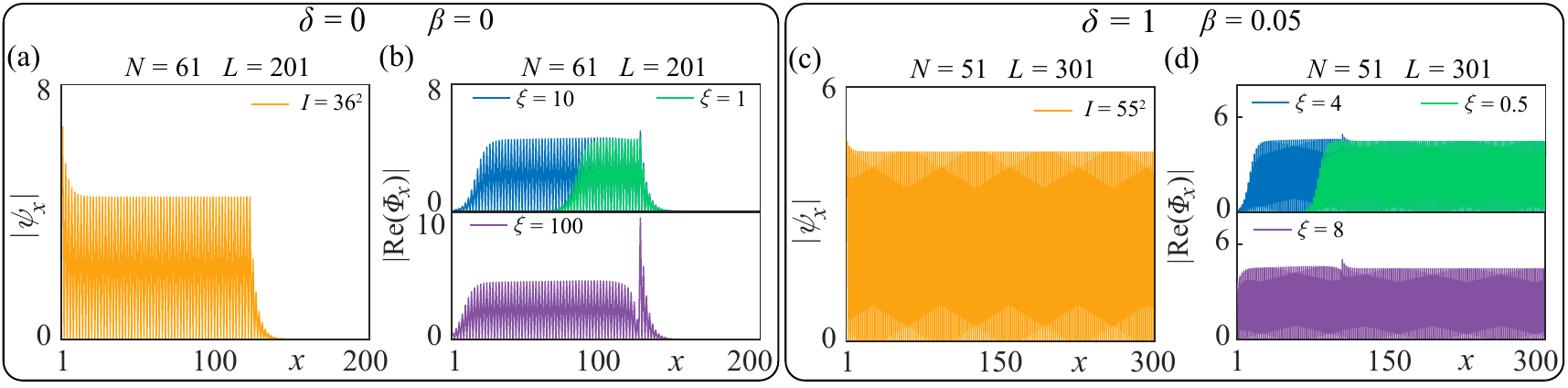}
	\caption{Hermitian case ($\delta = 0$ and $\beta = 0$): (a) static spatial distributions $\abs{\psi_x}$ of the extended TZM in a long lattice, and  (b) Wavefunction profiles $\abs{\text{Re}(\Phi_x)}$ of the evolved steady states under external pumping at a single site located at $2N+1$ with $\tilde{\omega}=0$, in accordance with the static results in (a). Non-Hermitian case 	($\delta = 1$ and $\beta = 0.05$): (c) static spatial distributions $\abs{\psi_x}$ of the extended TZM in a long lattice, and (d) Wavefunction profiles $\abs{\text{Re}(\Phi_x)}$ of the evolved steady states under external pumping at a single site located at $2N+1$ with $\tilde{\omega}=0$, in accordance with the static results in (c).	The parameters used are  $J = 1.5$, $\tau = t_\text{d} = 2.5$, $\tilde{t}_j=1.5$, and $\alpha = 0.05$, with $\tilde{\lambda}_j = 2.5$ for (a,b)  and $\tilde{\lambda}_j = 1.5$ for (c,d).}\label{FigS10}
\end{figure*}

\vspace{2mm}
\begin{center}\textbf{B. Achieving tailored wavefunction profiles with a long-range pattern through external pumping}\end{center}
\vspace{2mm}

In this discussion, we explore the ability to excite the targeted extended TZM with a long-range pattern, which could benefit a wide range of topological applications.

For the Hermitian case with $\delta = 0$ and $\beta = 0$, as shown in Fig.~\ref{FigS10}(a), we present a long-range uniform occupation of the extended TZM over $N=61$ unit cells in the static lattice. However, it is not possible to fully excite this long-range pattern through external pumping, as shown in Fig.~\ref{FigS10}(b). Even with a large pumping strength, the left-side lattice sites remain unexcited. In contrast, for the non-Hermitian case, such as for $\delta = 1$ and $\beta = 0.05$, as shown in Fig.~\ref{FigS10}(c,d), we can readily excite a long-range uniform occupation of the extended TZM over a much larger number of unit cells. This highlights the significant influence of the interplay between nonlinearity and non-Hermitian dynamics, which facilitates the excitation of extended modes over a much broader range compared to the Hermitian case.

\end{document}